\documentclass[prd, superscriptaddress, nofootinbib]{revtex4}

\usepackage{graphicx}
\usepackage{amssymb}
\usepackage{amsmath}
\usepackage{amsbsy}

\newcommand{\be}{\begin{equation}}
\newcommand{\ee}{\end{equation}}
\newcommand{\ba}{\begin{array}}
\newcommand{\ea}{\end{array}}
\newcommand{\bdm}{\begin{displaymath}}
\newcommand{\edm}{\end{displaymath}}
\newcommand{\pa}[1]{\left(#1\right)}
\newcommand{\paq}[1]{\left[#1\right]}
\newcommand{\ds}{\displaystyle}

\begin{document} 

\title{Cosmological model selection from standard siren detections by
third generation gravitational wave obervatories}
\author{Josiel Mendon\c{c}a Soares de Souza}
\affiliation{Departamento de F\'\i sica Te\'orica e Experimental, Universidade Federal do Rio Grande do Norte, Natal-RN 59072-970, Brazil}
\author{Riccardo Sturani}
\affiliation{International Institute of Physics, Universidade Federal do Rio Grande do Norte, Campus Universitario, Lagoa Nova, Natal-RN 59078-970, Brazil}

\begin{abstract} 
The multi-messenger observation of GW170817 enabled the first historic
measurement of the Hubble constant via a \emph{standard siren},
so-called in analogy to standard candles that enabled the measurement
of the luminosity distance versus redshift relationship at small redshift.
In the next decades, third-generation observatories
are expected to detect hundreds to thousand of gravitational wave events
from compact binary coalescences with potentially a joint electromagnetic
counterpart.
In the present work, we show how future standard siren detections can be used
within the framework of Bayesian model selection to discriminate between
cosmological models differing by the parameterization of the late-time acceleration.
In particular, we found quantitative conditions for the standard $\Lambda$CDM model
to be favored with respect to other models with varying dark energy content,
by reducing the uncertainty in the gravitational determination of the luminosity
distance with respect to current expectations.
\end{abstract}

\maketitle

\section{Introduction}
\label{sec:intro}
Coalescing binary systems in which at least one of the component is a neutron
star have long been considered the most likely candidate for a
simultaneous source of gravitational and electromagnetic radiation and indeed
GW170817, GRB170817A and SSS17a/AT 2017gfo \cite{TheLIGOScientific:2017qsa,GBM:2017lvd}
represented the first historic multi-messenger detection
involving gravitational waves and it was originated by the coalescence of two
neutron stars.
As first suggested in \cite{Schutz:1986gp}, see also \cite{Holz:2005df}, such coincidence detection
and the subsequent localization of the host galaxy can
be used to determine the Hubble constant $H_0$ by short-circuiting
the luminosity distance estimated via the gravitational channel and the
redshift measured electromagnetically.

The characteristic chirping gravitational wave signal of coalescing binary
enables an accurate determination of the intrinsic parameters of the source
(intrinsic luminosity) which combined with measured signal amplitude
(apparent luminosity) enable a determination  of the
luminosity distance, hence the name of \emph{standard sirens} for coalescing
binaries, in analogy to standard candles, the supernovae type Ia that
enabled the measurement of the Hubble constant and gave convincing
evidence of the late time
cosmological acceleration, see \cite{Riess:2016jrr} for recent observations.
 
The source redshift affects the gravitational wave signal degenerately
with the binary constituent intrinsic masses, making impossible its determination
from the gravitational signal alone. However
this degeneracy is not perfect and different approaches have been tried to
obtain a luminosity distance versus redshift relationship with
\cite{Messenger:2011gi}
and without electromagnetic counterparts, see \cite{DelPozzo:2011yh}.

Third-generation detectors like Einstein Telescope (ET) \cite{Punturo:2010zz}
and Cosmic Explorer \cite{Evans:2016mbw} are planned earth-based gravitational
wave detectors building on and improving the technology and
sensitivity of currently operating
detectors Advanced LIGO \cite{Harry:2010zz} and Advanced Virgo \cite{TheVirgo:2014hva} to reach sources up to redshift $z\sim $ few.
Given present rate estimates of binary neutron star coalescences of 
\cite{TheLIGOScientific:2017qsa} of $O(10^4)$Gpc$^{-3}$year$^{-1}$
detection rates of $O(10^3)$ events per year or larger are expected,
with large uncertainties due to the largely unknown
source distribution with redshift.
However, not all gravitational
  detections of binary system coalescences involving neutron stars are expected
  to be detected also electromagnetically, a reasonably optimistic
  expectation for electromagnetically bright standard sirens corresponding
  to few dozens of events per year \cite{Belgacem:2019tbw}.

With such a plethora of future data, we want to test the late-time dynamics
of the Universe via the redshift versus luminosity distance relationship.
At the moment the standard $\Lambda$CDM cosmological model
accommodates the observed late time acceleration ($z\lesssim 5$) as
well as other observations at larger redshift, however with a small
but significant tension in the value of the Hubble-Lema\^\i tre constant,
which is estimated to be $73.48\pm 1.66$ and $67.66\pm 0.42$km s$^{-1}$ Mpc$^{-1}$ respectively by standard candles \cite{Riess_last} 
and by Cosmic Microwave Background -based \cite{Aghanim:2018eyx} measures.
See also \cite{Birrer:2018vtm,refId0} for alternative determinations of $H_0$.

While at the moment no conclusive explanation of the discrepancy is
  available, it seems that early and late Universe determinations are at tension
  \cite{Verde:2019ivm}, leaving open the possibility that the same model may
  not fit precise observations from widely different cosmic epochs.

The \emph{dark energy} component required to explain current cosmological
acceleration is parameterized in the $\Lambda$CDM model by a bare cosmological
constant: a perfect fluid with negative pressure equal in modulus to
its energy density. While $\Lambda$CDM represents the standard
  cosmological model, the theoretical origin of the cosmological constant is
  completely unknown and as it stands, it is a phenomenological parameter whose
  value cannot be predicted and is uniquely determined by data fitting.
Indeed other phenomenological parameterizations are possible, involving an equal
or larger number of parameters than in the $\Lambda$CDM case, and the goal
of the present paper is to test if standard sirens detected by third-generation gravitational detectors can discriminate among different dark energy 
models, with possibly different number of parameters, giving a new handle to
solve the long-standing puzzle of what is the origin of the late time
cosmological acceleration.

We emphasize that the focus of the present work is \emph{not on parameter determination} with standard
sirens, that has been pursued in several recent publications, see e.g. \cite{Dalal:2006qt,Sathyaprakash:2009xt,Cai:2016sby,Zhang:2018byx,Chen:2017rfc}, but
rather \emph{model discrimination}, see \cite{Camera:2013xfa,Nishizawa:2017nef,Arai:2017hxj,Nishizawa:2019rra} for model comparison in specifically parameterized
non-GR models, and also \cite{Cai:2017yww} for a model-independent attempt to
reconstruct the distance versus redshift relationship via Gaussian process
methods with simulated LISA data.

However, current data are inconclusive to rule in or out a non-constant
dark energy, see e.g.~\cite{Abbott:2018wog,Aghanim:2018eyx}, and there are
indications that the variation with the redshift of dark energy may still be
compatible with zero once thousands of gravitational
waves will be detected, see e.g. \cite{Taylor:2012db,Belgacem:2019tbw}.
Future observations on the electromagnetic side, like the ones from
  Euclid \cite{Blanchard:2019oqi} up to redshift $z\lesssim 3$, are expected
  to reduce error bars on $H_0$ to per-mille level and measure
  $\Lambda$CDM deviation parameters $w_{0,a}$ with respectively $10^{-2},10^{-1}$
  precision and
  comparable precision is expected by combining third-generation
  gravitational wave detectors with other observations like cosmic
  microwave background and baryon acoustic oscillations \cite{Belgacem:2019tbw}.
  However, the power to rule in or out the $\Lambda$CDM model
  will not be related to non-zero of $w_{0,a}$ parameters, but also
  on the availability of models better describing the data,
  highlighting the importance to make data-driven forecast of different
  cosmological model comparisons.

In the present work, we take a new road: we use the Bayesian model selection
framework to compare pairs of different cosmological models, hence
we compute the \emph{evidence} of each model to rank them.

The outline of the paper is as follows: in sec.~\ref{sec:method}
we detail the method used to simulate data and rank models, in
sec.~\ref{sec:result} the results of Bayesian evidence computation are reported
and finally we conclude in sec.~\ref{sec:conclusion}.

\section{Method}
\label{sec:method}
The Hubble-Lema\^\i tre law relating redshift $z$ and luminosity distance $d_L$ via
the Hubble constant $H_0$ (we use natural units with $c=1$)
\be
d_L H_0=z+O(z^2)\,,
\ee
can be interpreted within the standard cosmological model based
on General Relativity as the first-order expansion of a more
general relationship between $z$ and $d_L$:
\be
d_L=\frac{1+z}{H_0}\int_0^z\frac{dz'}{E(z')}
\ee
where the inverse of the integrand
\be	
E(z) \equiv \sqrt{\Omega_{0M}(1+z)^3 + \Omega_{0R}(1+z)^4 + \Omega_{0DE}(1+z)^{3(1+w_{DE})}}
\ee
is expressed in terms of the normalized present energy densities $\Omega_{0X}$
in generic species $X$
\be
\Omega_{0X}\equiv \frac{8\pi G_N}{3H_0^2}\rho_{0X}\,,
\ee
with $\sum_X\Omega_{0X}=1$.
The equation of state relating pressure $p_X$ and energy density $\rho_X$ of each
species is assumed $p_X=w_X\rho_X$, implying
$\Omega_X\propto a^{-3(1+w_X)}$ (when no inter-species interactions are present)
and we have assumed that the only species
present in the Universe are non-relativistic matter $w_m=0$, radiation $w_r=1/3$
and dark energy with $w_{DE}$ that in the case of  cosmological constant becomes
$w_\Lambda=-1$.

Besides $\Lambda$CDM, we will use \emph{three} additional parameterizations of the dark energy in the rest of the paper:
\begin{itemize}
\item Model $w$CDM, with dark energy free parameter $w_{DE}=p_{DE}/\rho_{DE}$ constant in time.
\item Model $w_0w_a$CDM, with dark energy $w_{DE}=p_{DE}/\rho_{DE}=w_0+w_az/(1+z)$,
  with both $w_0$ and $w_a$ constant free parameters, as suggested in
  \cite{Chevallier:2000qy,Linder:2002et}, for which one has
  $\rho_{DE}(z)=\rho_{DE}(0)\pa{1+z}^{3\pa{1+w_0+w_a}}e^{-3w_a\frac z{1+z}}$.
\item The \emph{non-local massive gravity} model, henceforth \emph{massG},
described in \cite{Belgacem:2017cqo},
whose modified dynamics results in an identical luminosity distance for
electromagnetic waves as in General Relativity and in a modified one for
gravitational waves $d_L^{mG-gw}$ which can be phenomenologically
parameterised as
\be
\frac{d_L^{mG-gw}(z)}{d_L^{em}}=\Xi_0+\frac{1-\Xi_0}{(1+z)^n}\,,
\ee
with $n=5/2$ and $\Xi_0=0.97$ fixed, and two parameters
to fit to data which are $H_0$ and $\Omega_m$ as for $\Lambda$CDM
\cite{Belgacem:2018lbp}.
\end{itemize}

The $\Lambda$CDM, $w$CDM, $w_0w_a$CDM have the feature of being \emph{nested},
i.e. one can go from the more complex to the simplest by fixing one or
more parameters to specific values.
On the other hand, the non-local \emph{massG} model gives a
different description of late time Universe dynamics, still consistent with the data.
Since the fundamental origin of the cosmic acceleration is presently unknown,
it may well be that $\Lambda$CDM or its $w$-variants will not be able to
match at all values of redshift the dynamics resulting from the fundamental underlying
cosmological theory, hence we find it useful to adopt the \emph{massG}
model as a different, \emph{toy} model to
extract our simulated data from, to verify how different nested
parameterisations perform on data from a model none of them can match exactly.
We report in \ref{app:RR} our results for comparison of $\Lambda$CDM versus \emph{massG} which are in agreement with
  the ones of \cite{Belgacem:2018lbp}.

\subsection{Nested Models treatment: toy example}
In the case of nested models, the more general model always gives a better
fit by construction, but it can be disfavored as dictated by the \emph{Occam razor}
in case it uses unnecessary extra parameters. 
Let us see how nested model selection works in a simplified example that be
fully treated analytically \cite{Trotta:2008qt}.
The simplest case of 2 nested models consists of model ${\cal M}_0$ having no free parameter and ${\cal M}_1$ having one free parameter, say $\theta$, with ${\cal M}_1$ reducing to
${\cal M}_0$ for $\theta\to 0$.

If ${\cal M}_0$ describes the distribution of a Gaussian variable $x$ centered
in 0, then experimental data $d\equiv\{x_1,\ldots,x_n\}$ of mean $\lambda$ and standard deviation
$\sigma$ should result in a likelihood $L({\cal M}_0,d)$, or a probability distribution for $\lambda$, $p(\lambda|{\cal M}_0)$, given by
\bdm
p(\lambda|{\cal M}_0)=\frac 1{\sqrt{2\pi \sigma^2}}e^{-\lambda^2/(2\sigma^2)}\,.
\edm
${\cal M}_1$ on the other hand predicts that the measure of $x$ should be
described by a Gaussian centred in $\theta$, with $\theta$ a \emph{free}
parameter, to which an  $\emph{a priori}$ knowledge could be applied: e.g. we
assume it is distributed as a Gaussian with standard deviation $\Sigma$.
${\cal M}_1$ then predicts a probability distribution for $\lambda$ following a
$\theta$-dependent Gaussian distribution:
\bdm
p(\lambda|\theta,{\cal M}_1)=\frac 1{\sqrt{2\pi \sigma^2}}e^{-(\lambda-\theta)^2/(2\sigma^2)}\,.
\edm
According to standard Bayesian inference the probability distribution of
$\theta$ is given by
\be
\label{eq:pdf}
p(\theta|d,{\cal M}_1)=p(d|\theta,{\cal M}_1)\frac{p(\theta)}{p(d)}\propto p(d|\theta,{\cal M}_1)p(\theta)\,,
\ee
where $p(d|\theta,{\cal M}_1)$ is the likelihood of the data given parameter $\theta$
and model ${\cal M}_1$, $p(\theta)$ is the \emph{prior} on $\theta$
and in the last passage, we have dropped $p(d)$, which is uninteresting for
$p(\theta|d,{\cal M}_1)$ since it does not depend on $\theta$ and thus can be absorbed
in the normalization factor of $p(\theta|d,{\cal M}_1)$.

In this work we are interested in 
model comparison rather than parameter estimation, hence in comparing
$p(d|{\cal M}_0)$ to $p(d|{\cal M}_1)$ \emph{irrespectively} of the parameter values,
to know if data favor model ${\cal M}_0$ or ${\cal M}_1$.
This question can be addressed quantitatively by considering the ratio of the
\emph{evidences} $Z_{0,1}$
\be
\ba{l}
\ds\frac{Z_0}{Z_1}=\frac{p(d|{\cal M}_0)}{p(d|{\cal M}_1)}=\frac{p(d|{\cal M}_0)}{\int d\theta p(\theta)p(d|\theta,{\cal M}_1)}\\
\ds=\frac{(2\pi \sigma^2)^{-1/2}e^{-\lambda^2/(2\sigma^2)}}
{(2\pi \sigma^2)^{-1/2}(2\pi \Sigma^2)^{-1/2}\int d\theta e^{-\theta^2/(2\Sigma^2)}e^{-(\lambda-\theta)^2/(2\sigma^2)}}\,,
\ea
\ee
where as mentioned earlier we assumed a Gaussian prior on $\theta$
\be
p(\theta)=(2\pi \Sigma^2)^{-1/2}e^{-\theta^2/(2\Sigma^2)}\,.
\ee
After performing the integration in $\theta$ one gets
\be
\label{eq:an_lb}
\frac{Z_0}{Z_1}=e^{-\lambda^2/(2\sigma^2(1+\sigma^2/\Sigma^2))}
\sqrt{1+\frac{\Sigma^2}{\sigma^2}}\,,
\ee
showing that for $\lambda\ll \sigma$ ${\cal M}_1$ is disfavored for $\Sigma>\sigma$,
as expected by straightforward application of the Occam's razor,
and the models have similar evidences for $\Sigma\simeq \sigma$.\\
On the other hand for $\lambda\gg\sigma$ (and $\Sigma >\sigma$) $M_1$ quickly gains over $M_0$ despite the $\theta$ prior may disfavor large values
of $\theta$.
Note that ${\cal M}_1$ having more parameters and including ${\cal M}_0$ as
a particular case will always give a better fit to the data, but will not necessarily have better evidence.
In ~\ref{app:nested} we give a numerical check of eq.~(\ref{eq:an_lb}) in the specific problem studied in this paper.

\subsection{Merger rates and uncertainties}
Following \cite{Zhao:2010sz} we consider events for third-generation
gravitational wave detectors up to redshift
$z\sim 2$.
Actually gravitational signals could be seen up to much $z\sim 10$
\cite{Vitale:2016icu}, but we focus on a smaller range of redshift as
electromagnetic counterpart detections necessary for redshift determination
would be too difficult to observe from such large distances.
It is actually very optimistic to project that GW detections will have
electromagnetic counterparts even up to $z\sim 2$ \cite{Sathyaprakash:2019rom},
but as will see in sec.~\ref{sec:result}, discriminating power lies in small
error measurement at moderate redshift.

\begin{figure}
  \begin{center}
    \includegraphics[width=.6\linewidth]{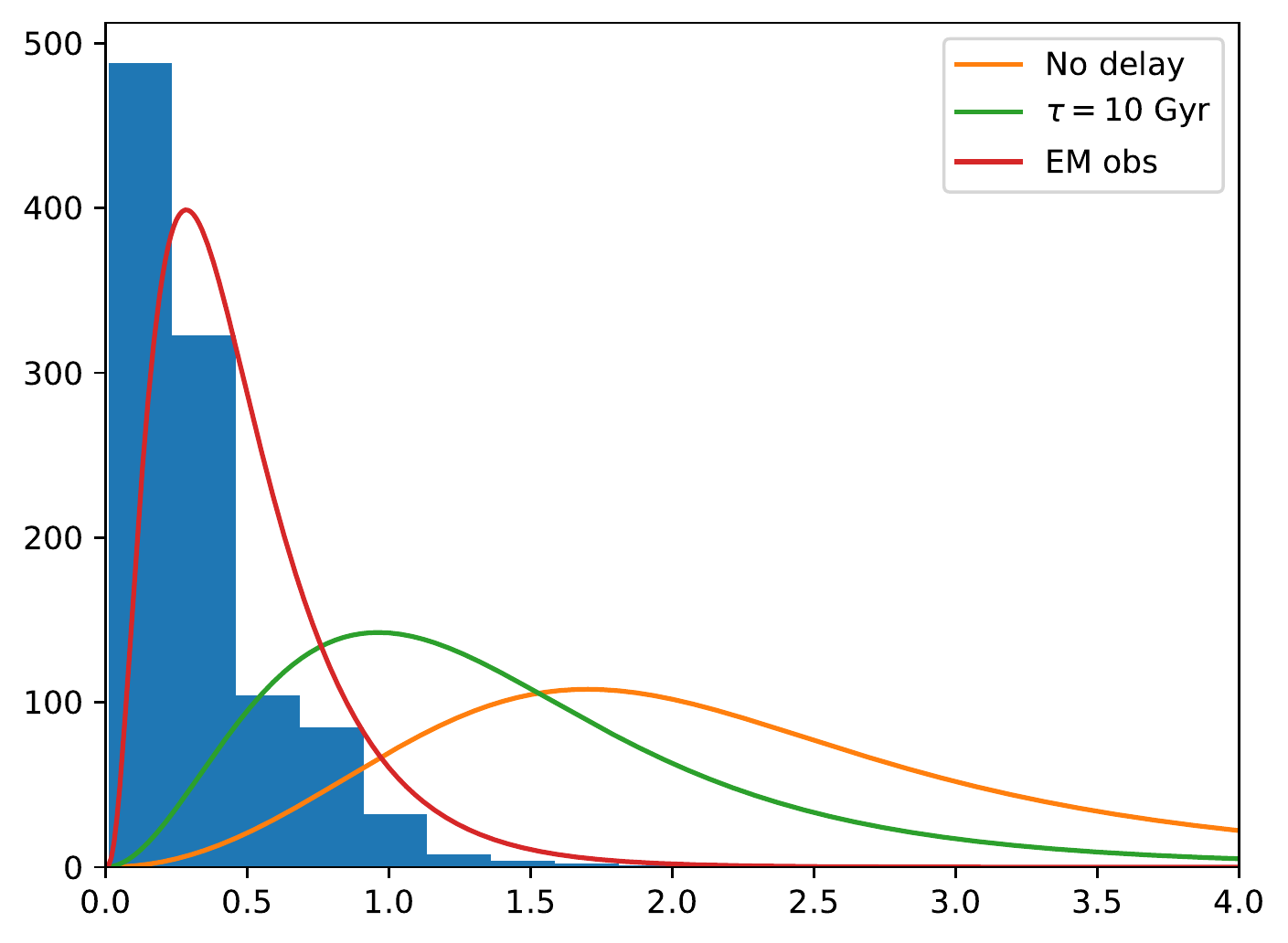}
    \caption{Histogram of redshift distribution of the 1048 supernovae used in
      \cite{Scolnic:2017caz} to measure the luminosity distance relationship (blue)
      compared with merger event distribution for different source distributions. Third-generation observatories can detect binary neutron stars up to $z\simeq 2$. Distributions have been normalised to have an equal number of events, 
      i.e. 1048.}
      \label{fig:sfr_merger}
    \end{center}
\end{figure}

To simulate the redshift distribution of signals we need the underlying
astrophysical distribution or merger rate which however is largely unknown.
To bypass such ignorance, we use three different curves for
  the distribution of merger events:
  \begin{enumerate}
  \item the \emph{star formation rate} per unit of comoving volume and redshift $\psi(z)$
    computed in \cite{Madau:2014bja}
    \be
    \label{eq:sfr}
    \psi(z)\propto\frac{\pa{1+z}^{2.7}}{1+\pa{\frac{1+z}{2.9}}^{5.6}}\,,
    \ee
  \item the same as above convolved with a stochastic delay between star formation
    and merger, assuming to follow a Poisson  distribution with characteristic
    delay $\tau=10$Gyr.
    Following \cite{Vitale:2018yhm,Safarzadeh:2019pis} one can find that the differential rate of
    mergers $R_m$ happening at redshift $z_m$ is given by
    \be
    R_m(z)=\frac{dN_m}{dt_odz}=\frac{dV_c}{dt_s}{\cal R}_m(z_m)\frac 1{1+z_m}\,,
    \ee
    in terms of merger rate ${\cal R}_m$ per unit of comoving volume $V_c$ and
    redshift $z$, and we have introduced the observer time $t_o$ related to source time $t_s$
    by $dt_o/dt_s=(1+z)$ and denoted with $N_m$ the number of mergers.
    ${\cal R}_m$ can in turn be modeled by convolving $\psi(z)$ with a stochastic \emph{delay}
    between star formation epoch characterized by redshift $z_{sf}$ and binary
    merger, which we assume for simplicity to be Poisson distributed, leading to
    \be
      \label{eq:poi_delay}
            {\cal R}_m(z_m)=\frac 1\tau\int_{z_m}^\infty dz_{sf} \frac{dt}{dz_{sf}}
            \psi(z_{sf})\exp\paq{-\frac{t(z_{sf})-t(z_m)}\tau}\,.
    \ee
  \item Not all binary coalescence involving neutron stars can be observed
    electromagnetically, hence a more realistic distribution \emph{EMobs} for standard
    sirens detectable in both the gravitational and electromagnetic channel
    can be parameterized as follows
    \be
    \label{eq:rateobs}
    {\cal R}_{EMobs}\propto\frac{z^3}{1+\exp\pa{10.7z^{0.6}}}\,,
    \ee
    obtained by a numerical fit to the distribution presented in
    \cite{Belgacem:2019tbw}.
\end{enumerate}
Fig.~\ref{fig:sfr_merger} shows the resulting normalized merger rate
of electromagnetically bright binary neutron star coalescence described above,
with a histogram of the redshifts of the supernova data \cite{Scolnic:2017caz,Meacher:2015iua} for comparison.

\begin{figure}
  \begin{center}
    \includegraphics[width=.6\linewidth]{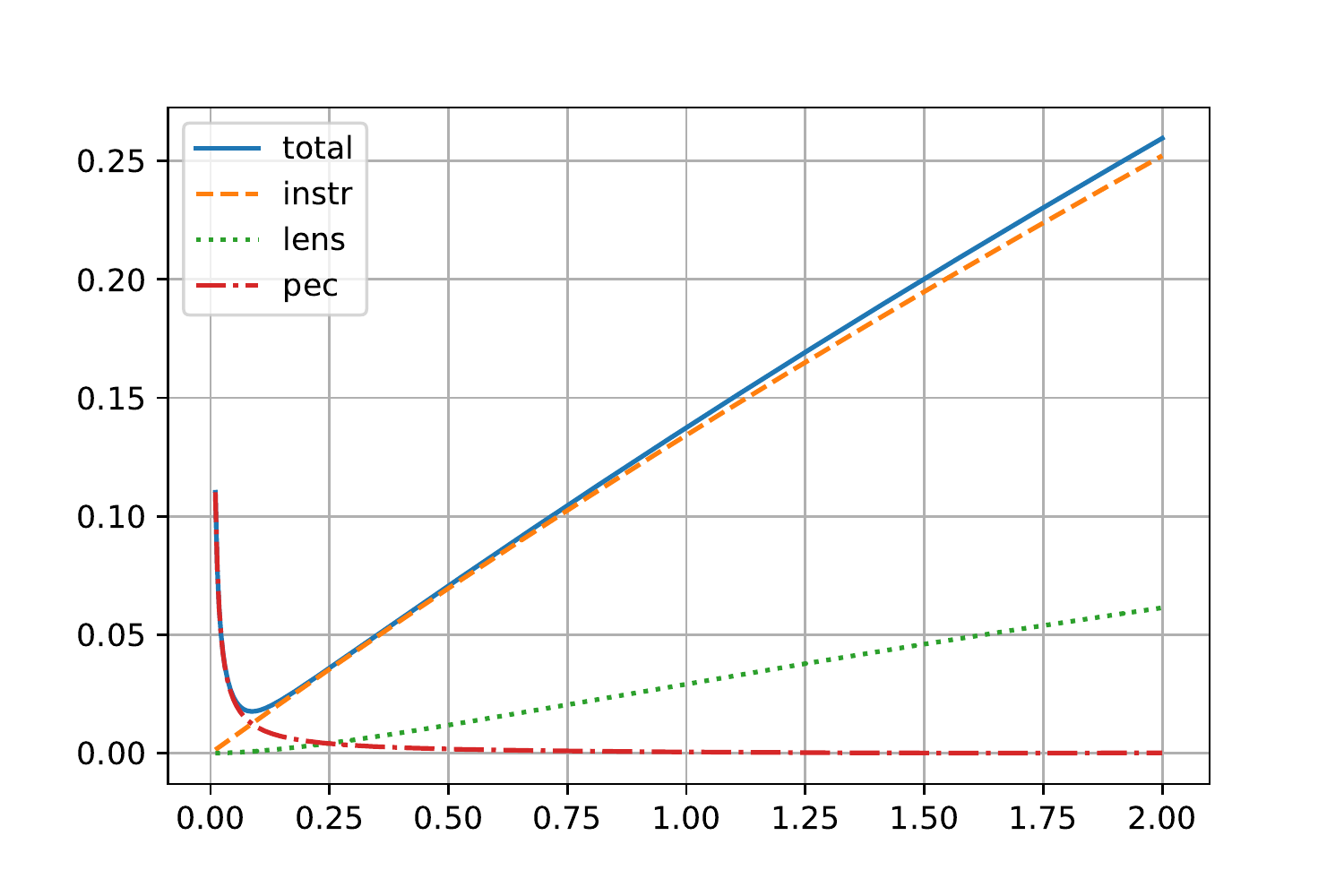}
    \caption{Uncertainty budget in the determination of the luminosity distance by
      third-generation detector like ET \cite{Zhao:2010sz,Hirata:2010ba,Gordon:2007zw} as a function of redshift,
      see eq.~(\ref{eq:errdL}). Instrumental effects give the leading uncertainty
      for all redshift of interest.}
      \label{fig:err_dL}
  \end{center}
\end{figure}

We assume that redshifts can be determined with negligible uncertainty
in the presence of an electromagnetic counterpart (see \cite{DelPozzo:2011yh,Abbott:2019yzh}
for measuring the Hubble constant $H_0$
\emph{without} electromagnetic counterpart) and the uncertainty
in $d_L$ has usually two main contributions: an instrumental one
intrinsic to GW observatories here denoted as $(\Delta d_L)_{inst}$ \cite{Zhao:2010sz}, and
another one $(\Delta d_L)_{lens}$ due to lensing, see e.g. \cite{Bonvin:2005ps,Hirata:2010ba}, see also \cite{Bonvin:2016qxr} for an additional source of bias in intrinsic parameters measure by GW detections).
Adding the sub-leading source of error due to peculiar velocities \cite{Gordon:2007zw} one has:
\be
\label{eq:dL_err}
\frac{\Delta d_L(z)}{d_L(z)} = \paq{
\left(\frac{\Delta d_L(z)}{d_L(z)}\right)_{inst}^2+
\left(\frac{\Delta d_L(z)}{d_L(z)}\right)_{lens}^2+
\left(\frac{\Delta d_L(z)}{d_L(z)}\right)_{pec}^2
}^{1/2}\,,
\ee
with
\be
\label{eq:errdL}
\ba{rcl}
\ds\left(\frac{\Delta d_L(z)}{d_L(z)}\right)_{inst} &\approx&\ds
0.1449z - 0.0118z^2 + 0.0012z^3\,,\\
\ds\left(\frac{\Delta d_L(z)}{d_L(z)}\right)_{lens} &\approx&\ds
0.066 \paq{4 \pa{1-(1+z)^{-1/4}}}^{1.8}\,,\\
\ds\left(\frac{\Delta d_L(z)}{d_L(z)}\right)_{pec} &\approx&\ds
\left|1-\frac{\pa{1+z}^2}{H(z)d_L(z)}\right|\frac{\sigma_v}c\,,
\qquad \sigma_{v}=331{\rm km/sec}\,,
\ea
\ee
which are shown in fig.~\ref{fig:err_dL} (with $c$ being the speed of light).

Note that a reduction in luminosity distance error may come from
combining more than one third-generation detector, as shown in
\cite{Vitale:2018nif}, leading to an error lower than 10\% in the red-shift
range of our interest.
This is why we also consider in the next section the case in which
the distance error is reduced to the lensing limit,
which roughly amounts to 20\% of the total error budget displayed in
fig.~\ref{fig:err_dL}.\footnote{A direct measure of the mass
distribution on the line of sight, which may be available by the time
GW data are collected, can reduce the lensing part of the error budget
\cite{Tamanini:2016zlh}.}.
The last source of error in redshift determination in eq.~(\ref{eq:dL_err},\ref{eq:errdL}),
the one due to peculiar velocities,
is important only for $z\lesssim 0.1$, where a negligible
number of detections are present, see fig.~\ref{fig:cum_dist_injs},
hence it will be neglected in this work.

\section{Results}
 \label{sec:result}
\begin{figure}
  \begin{center}
    \includegraphics[width=.6\linewidth]{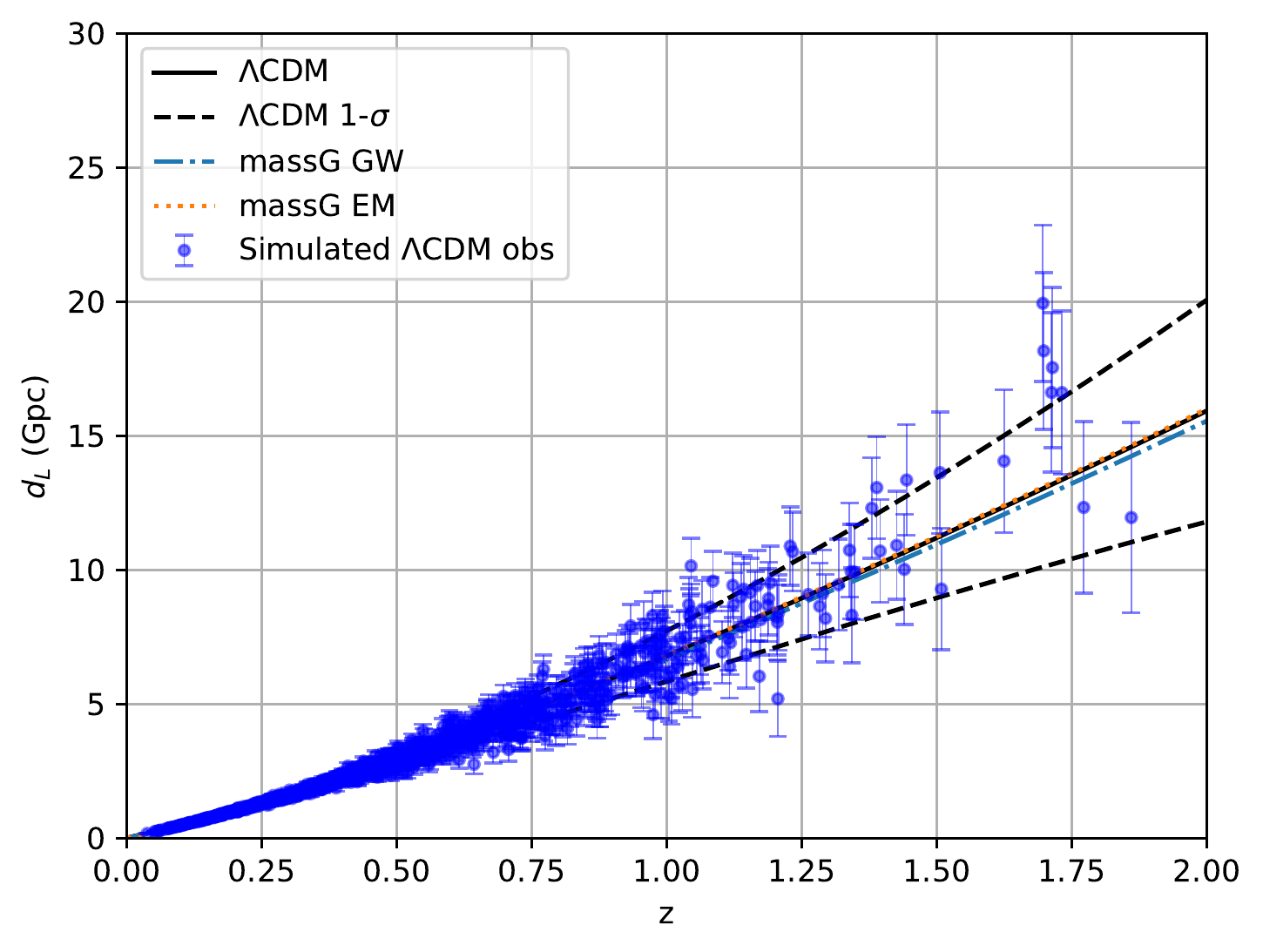}
    \caption{One thousand events simulated according to the \emph{EMobs} merger rate
      of eq.~(\ref{eq:rateobs}), with $1\sigma$ error band
      given by eq.~(\ref{eq:errdL}) \cite{Zhao:2010sz,Tamanini:2016zlh}.
      For reference, the luminosity distance versus redshift curve for $\Lambda$CDM
      and \emph{massG} models are also shown, for best fit parameters respectively
      given in \cite{Aghanim:2018eyx} and \cite{Dirian:2017pwp}.}
    \label{fig:injs}
  \end{center}
\end{figure}

\begin{figure}
  \begin{center}
    \includegraphics[width=.6\linewidth]{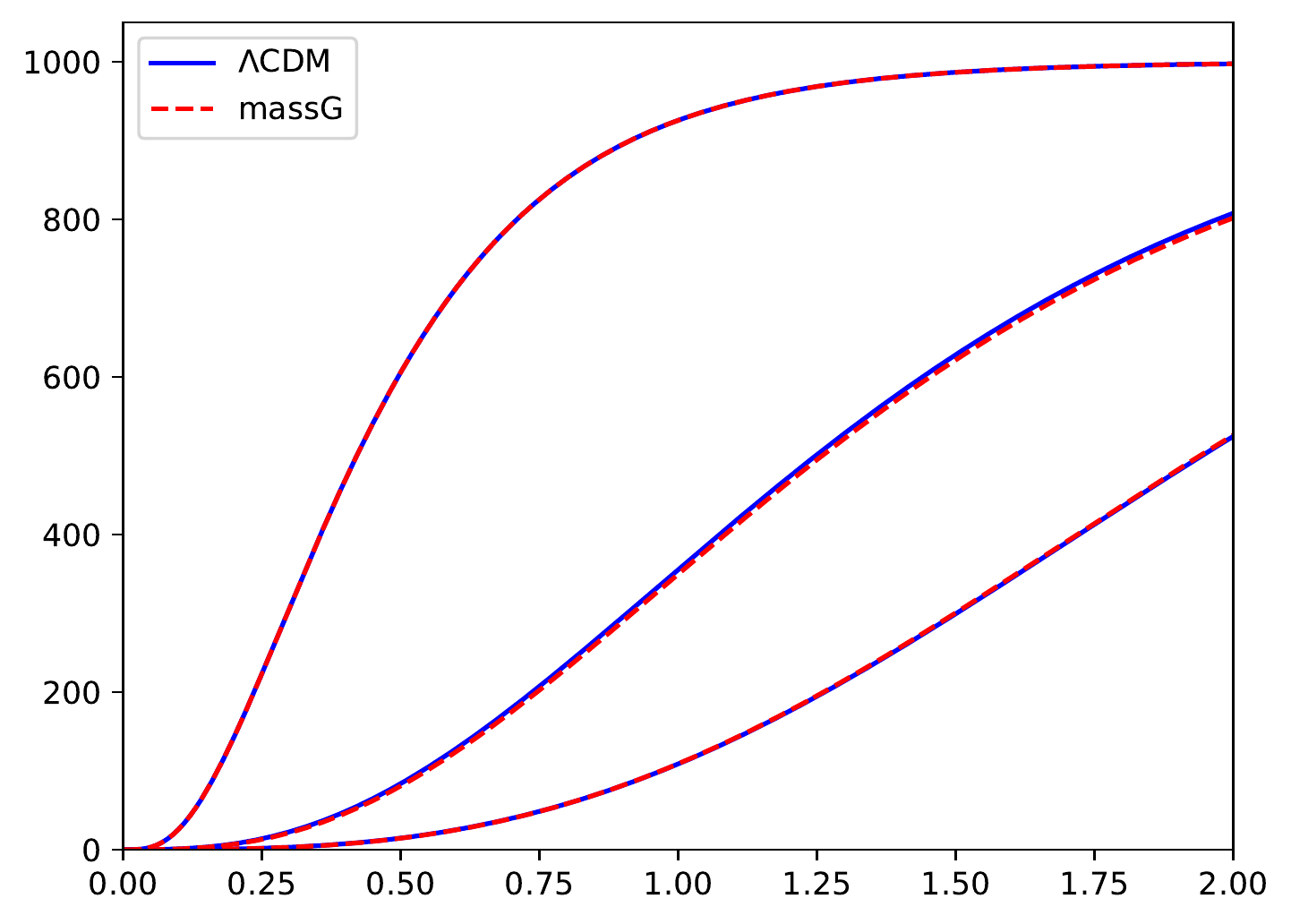}
    \caption{Cumulative distribution of injections for $\Lambda$CDM and the  
      \emph{massG} model\cite{Belgacem:2017cqo} as a function of redshift for the cases,
      from bottom to top, of no delay between star formation eq.~(\ref{eq:sfr})
      and binary mergers, $\tau=10$Gyr Poissonianly-distributed stochastic
      delay between star formation and mergers eq.~(\ref{eq:poi_delay}), and
      EM-observed distribution eq.~(\ref{eq:rateobs}).}
    \label{fig:cum_dist_injs}
  \end{center}
\end{figure}

This section reports the result obtained by simulating 1,000 detections of electromagnetically bright
coalescing binaries up to redshift $z=2$ with realistic distributions of
events and comparing phenomenological models for late time
cosmological acceleration within the Bayesian model selection framework.
We use \emph{two different} laws for relating the standard siren luminosity distance and redshift of the data:
\begin{enumerate}
  \item the \emph{standard $\Lambda$CDM} which we also use at recovery, with parameters $\Omega_m=0.3111$, $H_0=67.66\rm{km/s/Mpc}$,
$\Omega_\Lambda=1-\Omega_m$  taken from \cite{Aghanim:2018eyx},
  \item the \emph{non-local massive gravity} (\emph{massG}) model
\cite{Belgacem:2017cqo}, useful as a testing ground for different models
at recovery, none of which include the \emph{massG} model used in this
second set of injections. Note that
since the background evolution in this model is different than any of the $\Lambda$CDM
and $w$CDM, the best-fit background parameter value are slightly different
than in the previous case: $\Omega_m=0.2989$, $H_0=69.49\rm{km/s/Mpc}$ \cite{Dirian:2017pwp}.
\end{enumerate}
Each of the two sets of simulated data ($\Lambda$CDM and \emph{massG}) is
produced for \emph{three different distributions of merger events}:
\begin{enumerate}
 \item one following the \emph{star formation rate} proposed in
\cite{Madau:2014bja}, orange solid line in fig.~\ref{fig:sfr_merger},
 \item the second allowing a \emph{Poisson distributed delay}
between star formation and binary merger with average delay $\tau=10$ Gyr,
green solid line in fig.~\ref{fig:sfr_merger},
\item the last one following the realistic observed rate given in eq.~(\ref{eq:rateobs}),
  referred to as \emph{EMobs}, red solid line in fig.~\ref{fig:sfr_merger},
\end{enumerate}
resulting in \emph{six different} sets of injections.
Fig.~\ref{fig:injs} displays explicitly one of the six types of injections
we use, and in fig.~\ref{fig:cum_dist_injs} the cumulative distributions of the
1,000 injections are reported for the six cases,
showing little difference between the $\Lambda$CDM and $massG$ case, but a notable
difference among the three underlying cosmological distributions of mergers.

\begin{figure}
\begin{center}
\includegraphics[width=.48\linewidth]{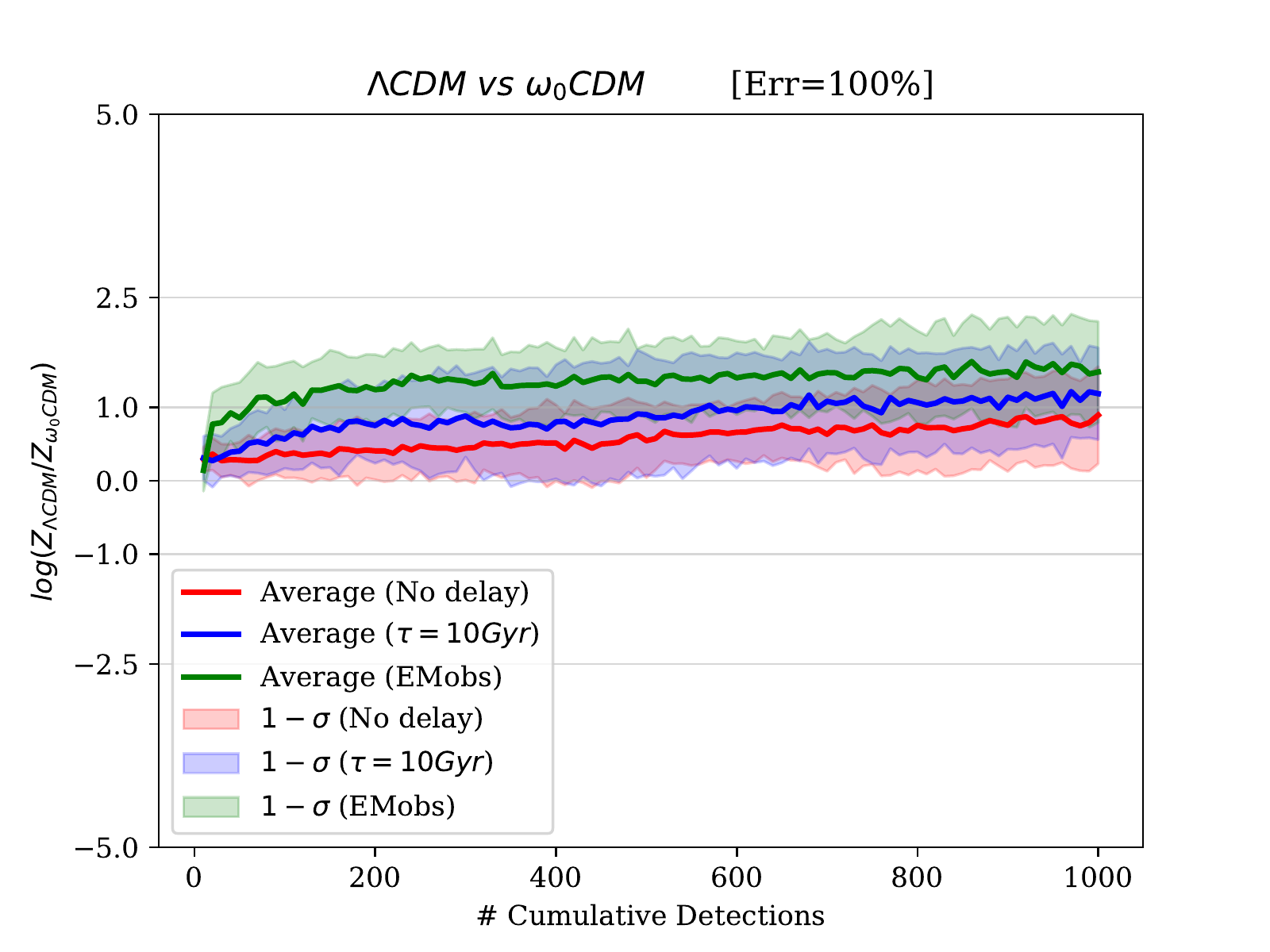}
\includegraphics[width=.48\linewidth]{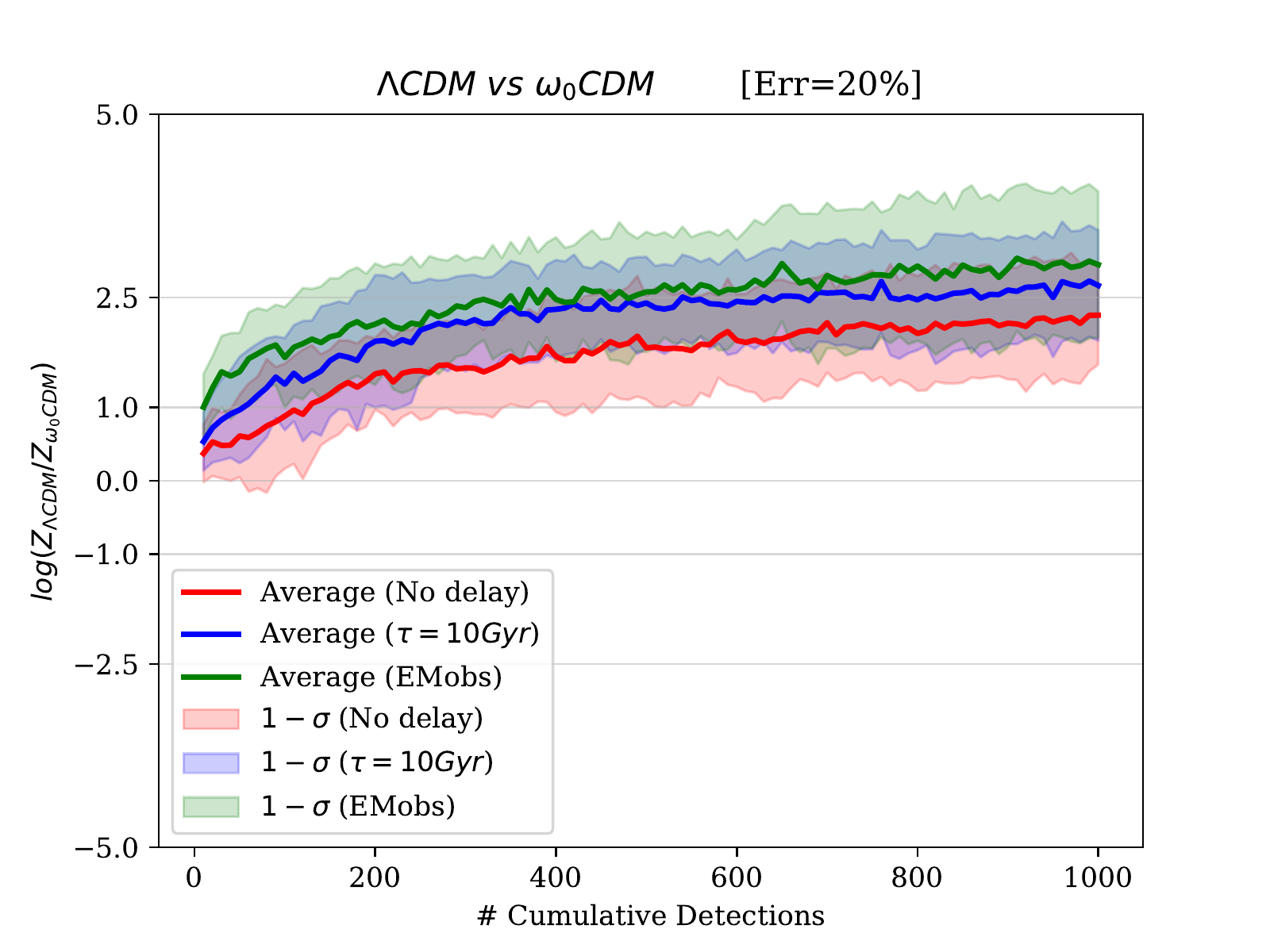}\\
\caption{Evidence for $\Lambda$CDM versus $w$CDM model with 1,000 events,
simulated according to the $\Lambda$CDM model
with standard uncertainty in the luminosity distance given by eq.~(\ref{eq:errdL}) (left) and uncertainty reduced to $20\%$ (right), 
with merger rate equal to star formation rate, convolved with a Poisson-distributed $\tau=10$ Gyr delay as in eq.~(\ref{eq:poi_delay})
between formation and merger, and according to \emph{EMobs} distribution eq.~(\ref{eq:rateobs}).
From 50 different injection realizations the shady regions corresponds to 1-$\sigma$ level around the thick line showing the average value.
Weak, moderate and strong evidence reference lines are also displayed
for $\ln Z_1/Z_2=\pm 1,\pm 2.5, \pm 5$) according to Jeffreys' scale.}
\label{fig:Lvsw0LCDMinjs}
\end{center}
\end{figure}

\begin{figure}
\begin{center}
\includegraphics[width=.48\linewidth]{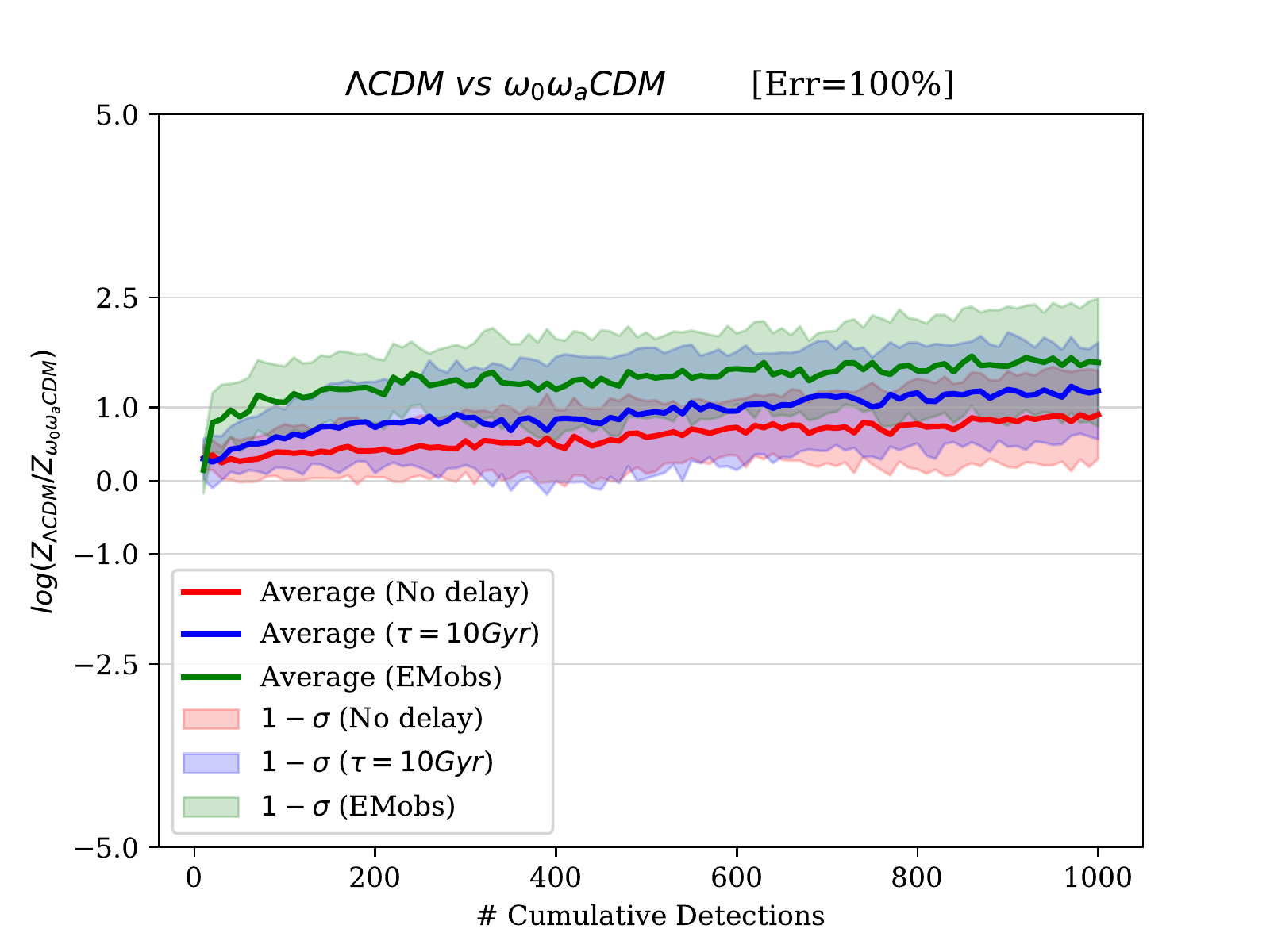}
\includegraphics[width=.48\linewidth]{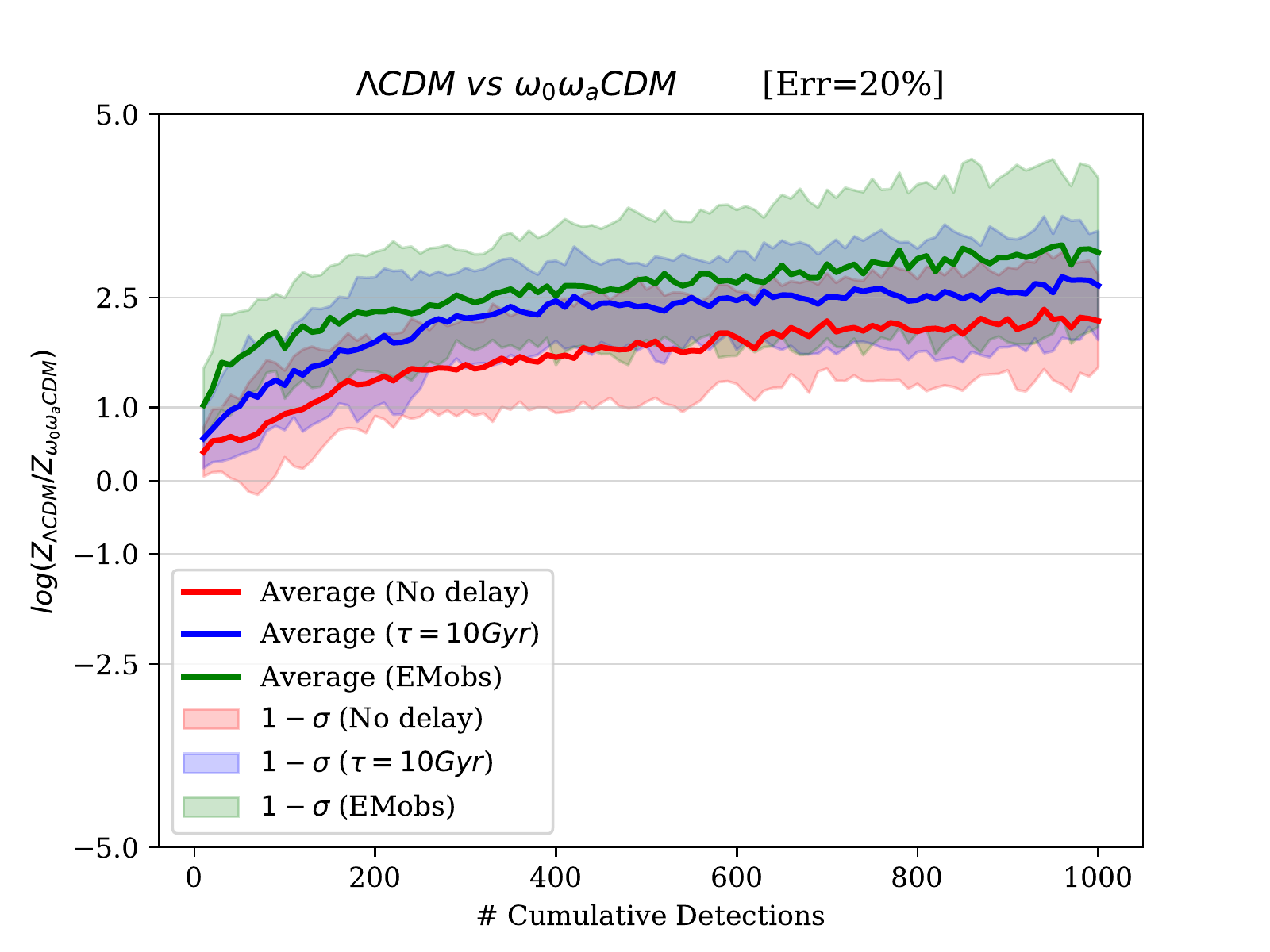}\\
\caption{Analog to fig.~\ref{fig:Lvsw0LCDMinjs} for $\Lambda$CDM versus
$w_0w_a$CDM model: simulated 1,000 events following $\Lambda$CDM model,
standard $d_L$ uncertainty (left) and reduced uncertainty, $20\%$ of
eq.~(\ref{eq:errdL}) (right),
with merger rate equal to star formation rate, convolved with a Poisson-distributed $\tau=10$ Gyr delay as in eq.~(\ref{eq:poi_delay})
between formation and merger, and according to \emph{EMobs} distribution eq.~(\ref{eq:rateobs}).
From 50 different injection realizations the shady regions corresponds to 1-$\sigma$ level around the thick line showing the average value.}
\label{fig:Lvsw0waLCDMinjs}
\end{center}
\end{figure}

\begin{figure}
\begin{center}
\includegraphics[width=.48\linewidth]{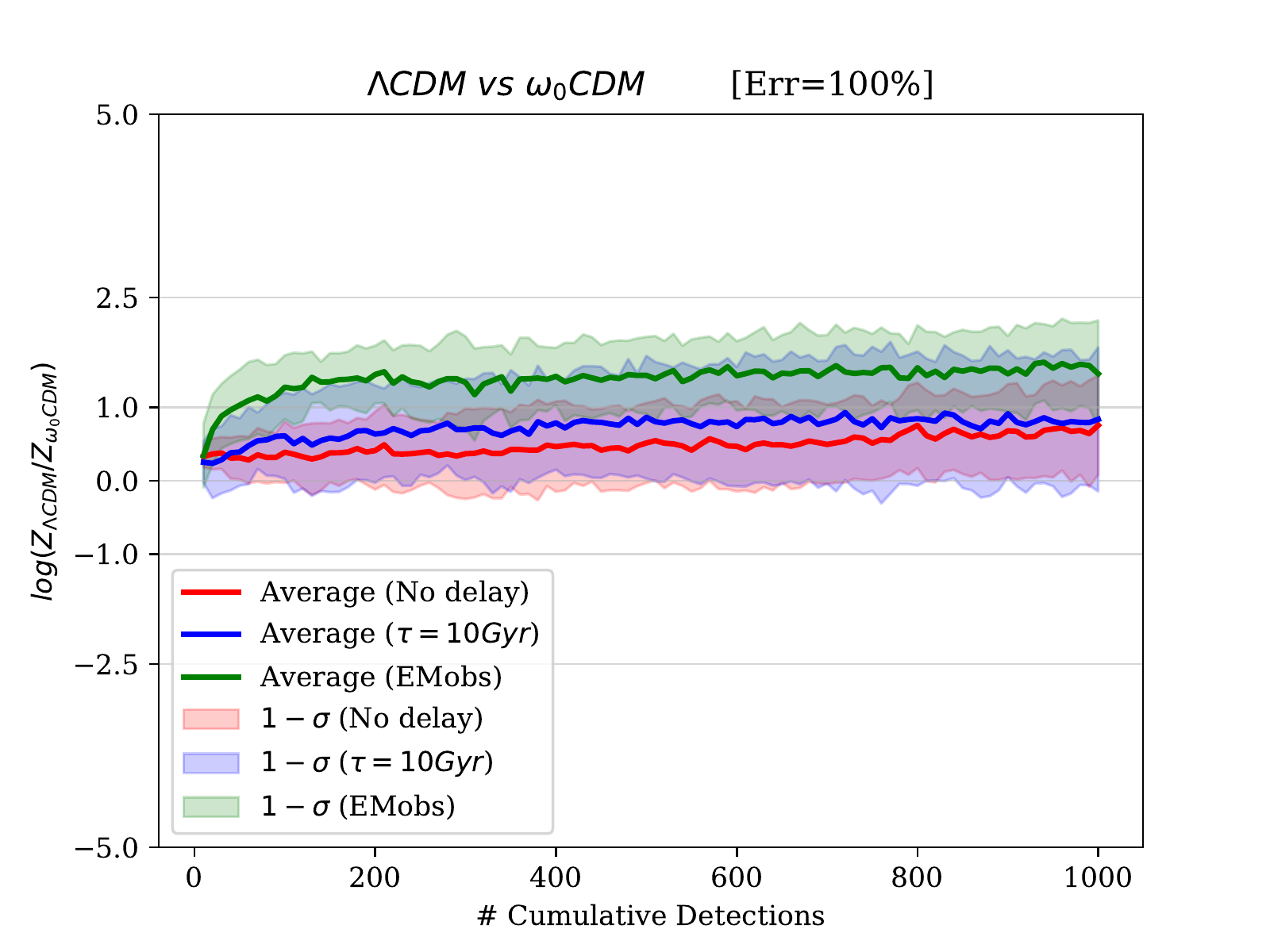}
\includegraphics[width=.48\linewidth]{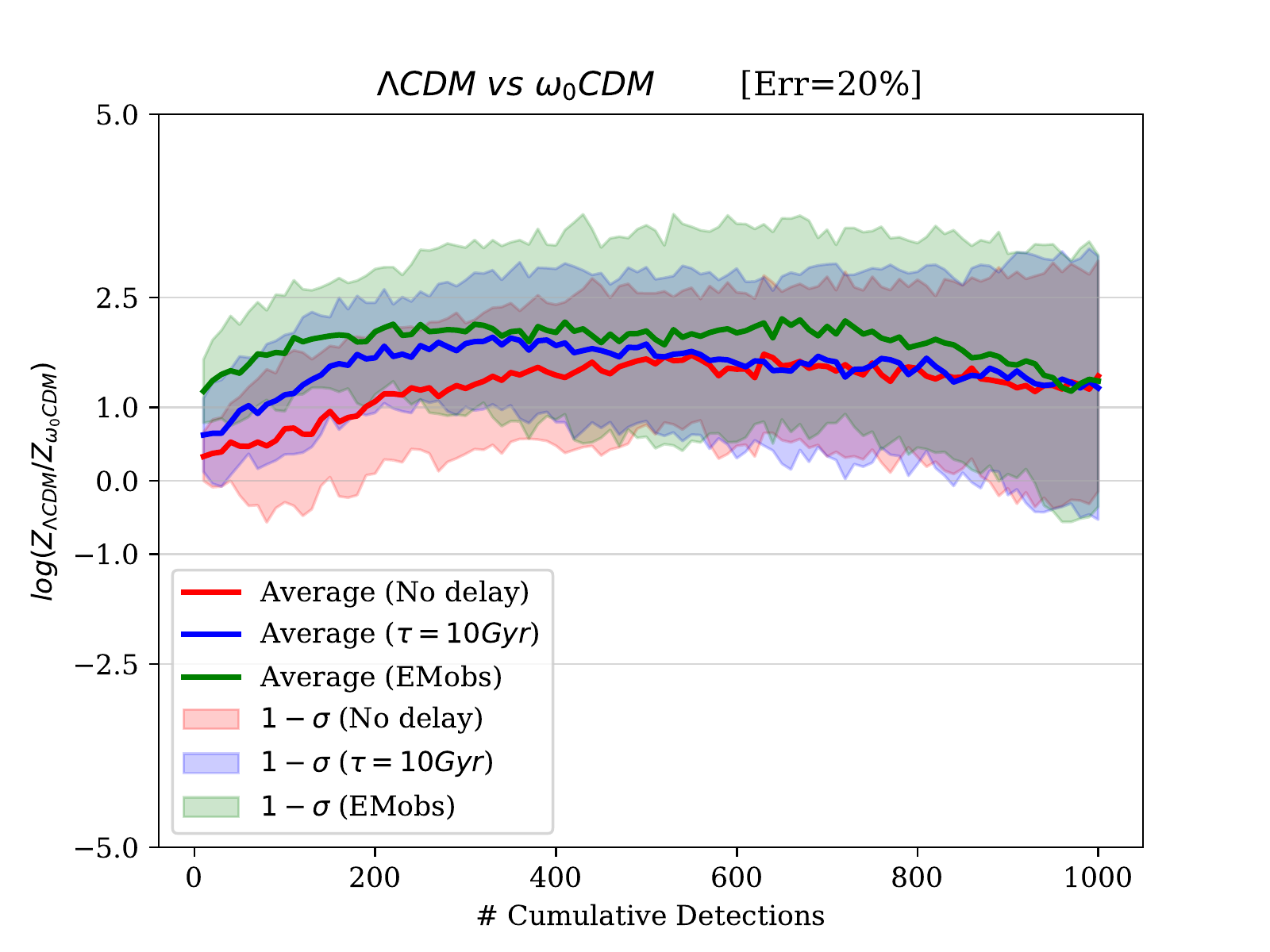}\\
\caption{Analog to fig.~\ref{fig:Lvsw0LCDMinjs} for simulated data following
the \emph{massG} model \cite{Belgacem:2017cqo}: evidence for $\Lambda$CDM versus $w$CDM
models with 1,000 events, standard $d_L$ uncertainty eq.~(\ref{eq:errdL}) (left) and uncertainty reduced to 20\% (right),
with merger rate equal to star formation rate, convolved with a Poisson-distributed $\tau=10$ Gyr delay as in eq.~(\ref{eq:poi_delay})
between formation and merger, and according to \emph{EMobs} distribution
eq.~(\ref{eq:rateobs}).
From 50 different injection realizations the shady regions corresponds to 1-$\sigma$ level around the thick line showing the average value.
Moderate and strong evidence reference lines ($\ln Z_1/Z_2=\pm 2.5,\pm 5$ according to Jeffreys' scale) are also shown for reference.}
\label{fig:Lvsw0MassGRinjs}
\end{center}
\end{figure}

\begin{figure}
\begin{center}
\includegraphics[width=.48\linewidth]{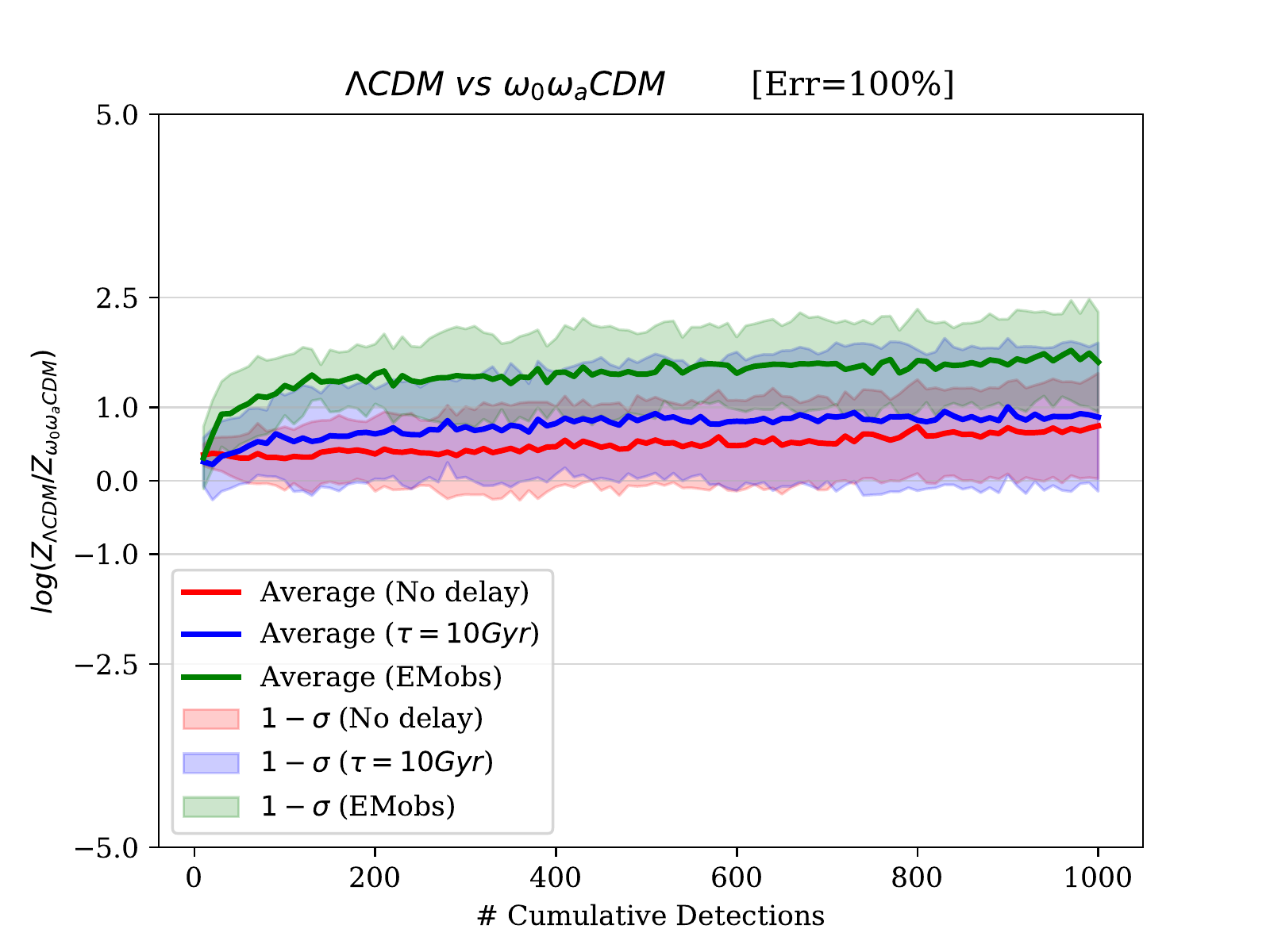}
\includegraphics[width=.48\linewidth]{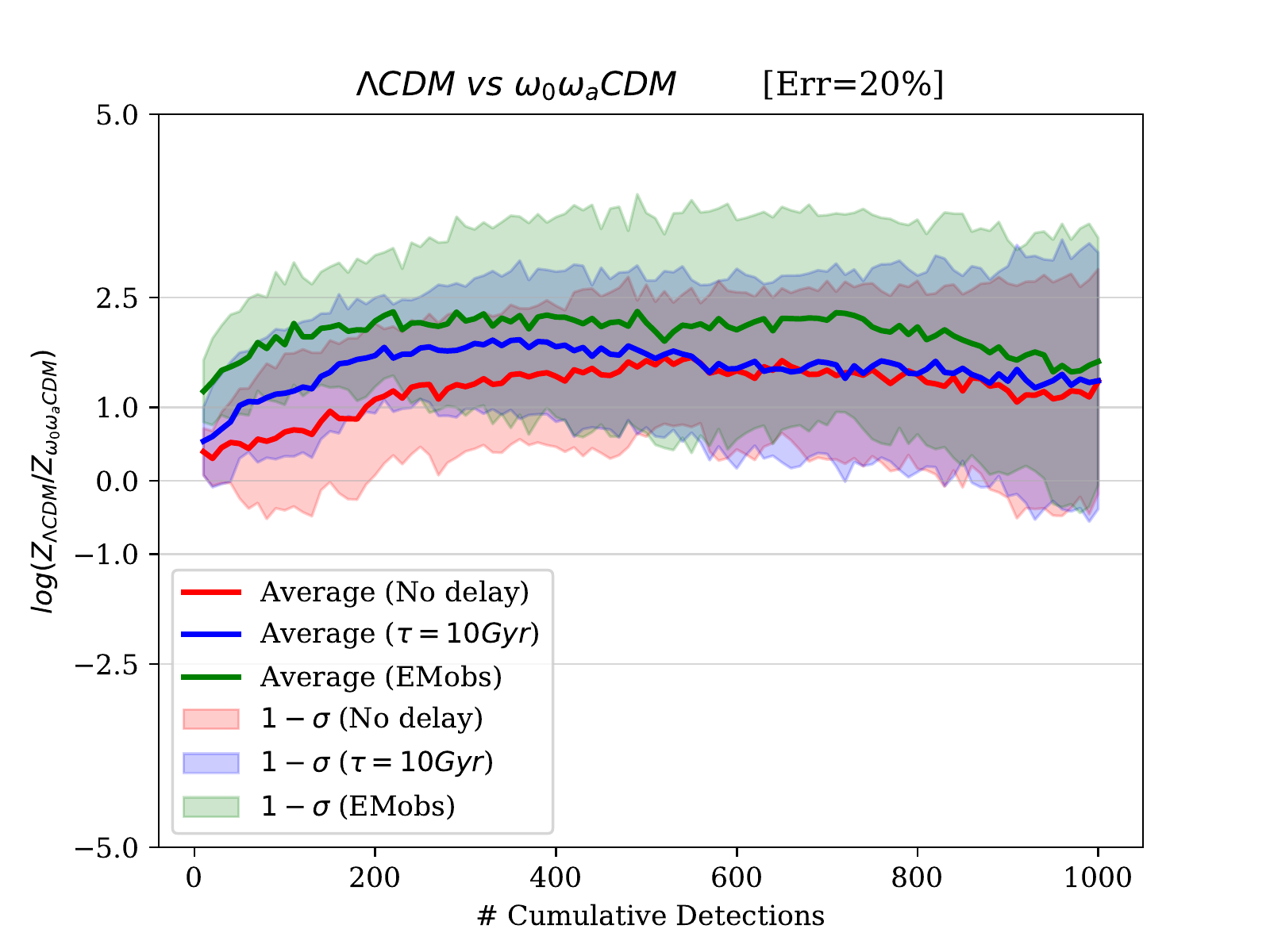}\\
\caption{Same as in fig.~\ref{fig:Lvsw0LCDMinjs} for simulated data following the
  \emph{massG} model\cite{Belgacem:2017cqo}: evidence for $\Lambda$CDM versus $w_0w_a$CDM
models with 1,000 events. Standard $d_L$ uncertainty (left) and
reduced uncertainty (right),
with merger rate equal to star formation rate, convolved with a Poisson-distributed $\tau=10$ Gyr delay as in eq.~(\ref{eq:poi_delay})
between formation and merger, and according to \emph{EMobs} distribution
eq.~(\ref{eq:rateobs}).
From 50 different injection realizations the shady regions corresponds to 1-$\sigma$ level around the thick line showing the average value.}
\label{fig:Lvsw0waMassGRinjs}
\end{center}
\end{figure}

We use these six type of simulated data for comparing $\Lambda$CDM versus
$w$CDM and $\Lambda$CDM versus $w_0w_a$CDM, with the results for model
comparisons displayed respectively in figs.~\ref{fig:Lvsw0LCDMinjs},\ref{fig:Lvsw0waLCDMinjs} for $\Lambda$CDM injections,
and in figs.~\ref{fig:Lvsw0MassGRinjs},\ref{fig:Lvsw0waMassGRinjs}
for $massG$ injections.
The 1,000 simulated detections are divided in 100 catalogs of 10 detections each,
for each graph 50 different injections realizations have been performed,
  shady regions representing 1-$\sigma$ level around the thick line showing
  the average value.

Results are obtained via the \emph{Nestle} implementation \cite{Mukherjee:2005wg}
of the nested sampling algorithm \cite{Skilling:2006gxv} using 200 live points.
Priors for all have been chosen flat in the intervals:
$H_0: [60,80]$ km/s/Mpc, $\Omega_m: [0.2,0.4]$, $w_0: [-2,0]$, $w_a: [-1,1]$.

Fig.~\ref{fig:Lvsw0LCDMinjs} shows the evidence comparison for $\Lambda$CDM
versus $w$CDM, with simulated data following the $\Lambda$CDM model
and taken from a distribution of merger events following the star formation rate of eq.~(\ref{eq:sfr}) \cite{Madau:2014bja},
the same star formation rate convoluted with a delay with a Poisson distribution
with average $\tau=10$Gyr, and EMobs rate (\ref{eq:rateobs}),
each of the three cases with luminosity distance uncertainties as in
eq.~(\ref{eq:dL_err}) and with uncertainties reduced to 20\% that value.

All evidence ratios correctly favor $\Lambda$CDM vs. $w$CDM, as
expected since injections follow the $\Lambda$CDM model, and
reducing observational uncertainty strengthens the discriminating
power of the Bayes model selection test, with the logarithm of evidence
ratio approaching the weak to moderate evidence line
$\ln Z_{\Lambda CDM}/Z_{wCDM}=2.5$, see \emph{Jeffreys' scale} \cite{jeffreys,Trotta:2008qt},
for the reduced error cases. Note also that having the detections
at lower redshifts, where the luminosity distance
measure error is smaller, gives more discriminating power.

Analogously in fig.\ref{fig:Lvsw0waLCDMinjs} $\Lambda$CDM is ranked against
$w_0w_a$CDM, i.e. against a model with two extra parameters that contains
$\Lambda$CDM for $w_0=-1$, $w_a=0$ for two sets of $d_L$ uncertainties and three
sets of signal distributions as above, again with the result
of the $\Lambda$CDM model being favored.
Again reducing the measurement errors and taking the detections at
relative lower redshifts have the effect of increasing
the discriminating power of the data, showing that event distribution with
redshift also plays an important role.

In figs.~\ref{fig:Lvsw0MassGRinjs},\ref{fig:Lvsw0waMassGRinjs} the same exercise
is replayed, with the difference of using injections belonging to the
\emph{massG} model \cite{Belgacem:2017cqo}, hence not being described by
any of the nested models used for recovery, which are again $\Lambda$CDM and 
$w$CDM in fig.~\ref{fig:Lvsw0MassGRinjs}, and $\Lambda$CDM and $w_0w_a$CDM in
\ref{fig:Lvsw0waMassGRinjs}.
In this case, in which none of the recovery models coincide with the one
used for injections, irrespectively of the assumptions on the error on the
luminosity distance and event distribution, the outcome of evidence
comparison is inconclusive, and specific realizations can give logarithm of
Bayes' factor ratios with both signs.
This indicates that future inconclusive results in comparing models may be due to the lack
of appropriate parameterization to describe data.

\ref{app:RR} reports the comparison between $\Lambda$CDM and \emph{massG} models
over simulations based on the \emph{massG} model.

\section{Conclusions}
\label{sec:conclusion}
Motivated by the advent of gravitational wave astronomy and by the first
measure of the Hubble constant via standard sirens,
we have performed a numerical exercise simulating future measurements of the
luminosity distance versus redshift relationship via combined detections
of gravitational and electromagnetic waves to test their power in
discriminating among cosmological models of late time acceleration.

There are two main conclusions that we can draw from our simulations.
The first is that the error in luminosity distance as estimated in eq.~(\ref{eq:dL_err})
should be decreased to enable model comparison with future gravitational wave
detectors.
This should be possible by correlating the output of several observatories
\cite{Vitale:2016icu}:
in particular, for the case of the Einstein Telescope combining several detectors
can lead to substantial improvement (a factor of few) in the luminosity
distance error with respect to a single detector.
However even reducing at maximum the detector intrinsic uncertainty, the lensing
limit becomes a limit factor for the measure of the luminosity distance, as
also underlined in \cite{Belgacem:2019tbw}.

The second main conclusions are that the intrinsic event distribution also
plays an important role: distributing events at smaller distances concentrate
events where observational error are smaller and hence convey more information,
even if different models tend all to reproduce the same dynamics
for small redshifts ($z\lesssim 1$) and disagree more at larger redshifts
($z\gtrsim 2$).

Note that the expected rate of standard siren accompanied by electromagnetic
  detections is subject to large uncertainties, but a reasonably optimistic
  expectation corresponds to few dozens of events per year,
  thus requiring several years to accumulate $O(10^3)$ detections
  \cite{Belgacem:2019tbw}.

Finally, we underline that in case the dynamics underlying observations are only
approximately described by models used to analyze data, 
different models may have comparable performance on data, leaving open
the search for a better model able to catch the right physics conveyed by the
observations.

We also note that while cosmological model discrimination at low redshift may be
  hard, future observations of gravitational standard sirens by LISA
  in conjunction with Quasar electromagnetic detections at $z\gtrsim 3$
  may be powerful for luminosity distance typical precision of the order of
  10\%, as demonstrated in \cite{Speri:2020hwc}, where $\Lambda$CDM can be
  discriminated against the alternative model originally proposed in
  \cite{Risaliti:2018reu} already with a handful of detections.\\
  On the other hand future joint analysis of electromagnetic and
  gravitational observations are expected to reach sub-percent precision
  in $\Lambda$CDM deviation parameters $w_{0,a}$, helping the quest for
  a more fundamental cosmological model.

\section*{Acknowledgments} 
The authors thank Luciano Casarini, Jailson Alcaniz, and Rodrigo de Hollanda for
discussions and the anonymous referee for useful suggestions.
This work is the based on the thesis prepared by JMSdS to obtain the master's
degree in physics at the Federal University of Rio Grande de Norte in Natal (Brazil).
The work of JMSdS has been partly financed by the Coordena\c{c}\~ao de
Aperfei\c{c}oamento de Pessoal de N\'\i vel Superior - Brasil (CAPES) -
Finance Code 001.
RS thanks CNPq for partial financial support.
This work was supported by ICTP-SAIFR FAPESP grant 2016/01343-7.
We  thank  the  High  Performance  Computing  Center  (NPAD)  at  UFRN  for
providing computational resources.

\appendix
\section{Comparison of nested models}
\label{app:nested}

\begin{figure}
  \begin{center}
    \includegraphics[width=.6\linewidth]{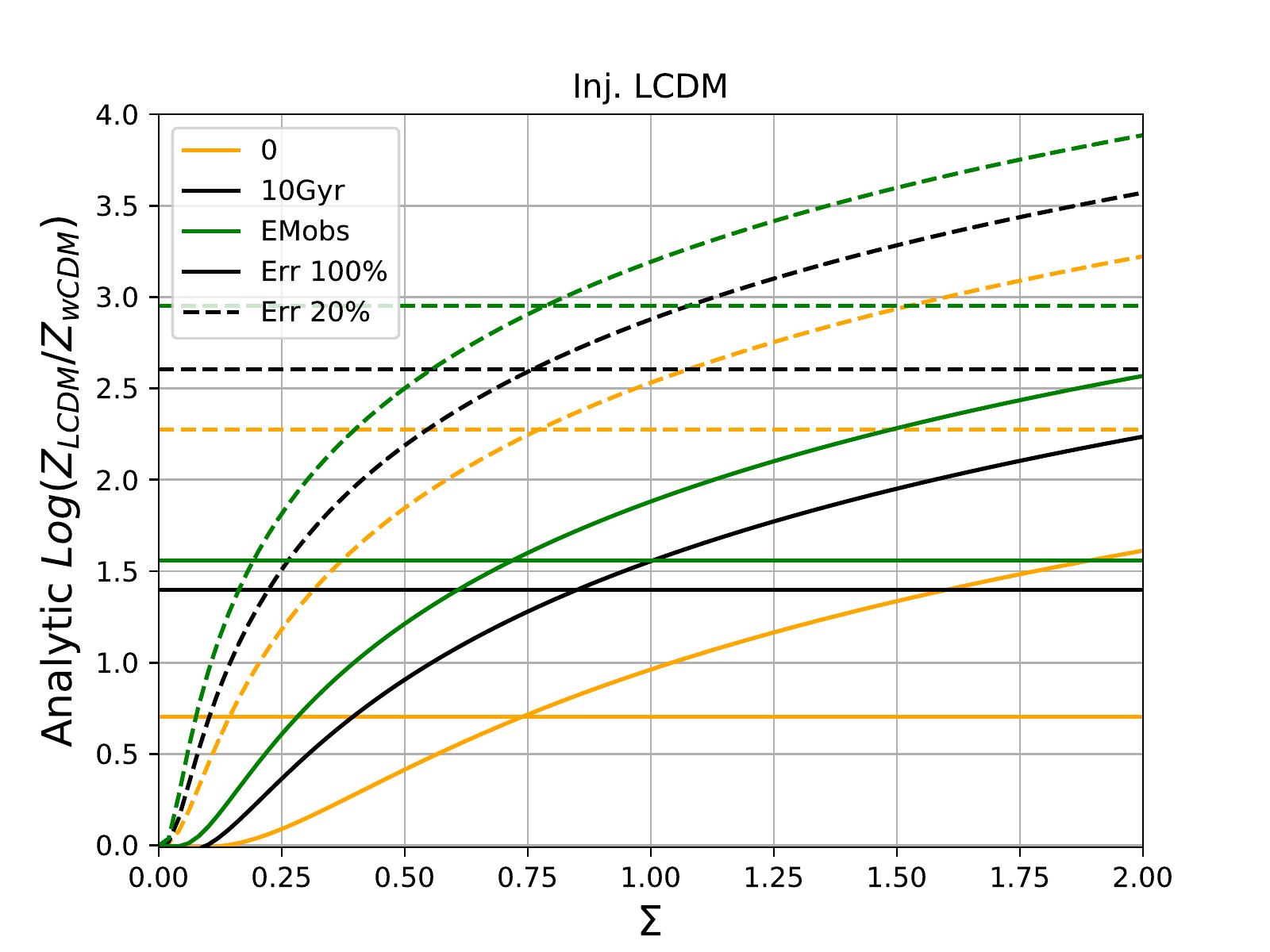}
    \caption{Comparison between analytic estimate of log Bayes ratio for
      $\Lambda$CDM vs. $w$CDM model from eq.~(\ref{eq:an_lb}) as a function
      of $w_0$ prior width $\Sigma$ compared with the values obtained numerically
      (horizontal lines) in the simulation described in this work, averaging over 50
      realisations of 1,000 injections.}
    \label{fig:lbnested}
\end{center}
\end{figure}

As additional material, we report in fig.~\ref{fig:lbnested}
the comparison of the logarithm of evidence ratios between nested
models $\Lambda$CDM and $w$CDM obtained numerically after averaging over 50 sets
of 1,000 $\Lambda$CDM injections each,
and the theoretical analytic estimate in eq.~(\ref{eq:an_lb}) as a function on $\Sigma$, the width of the prior on the
extra variable. We remind the reader that in our numerical analysis the prior over $w_0$
  is flat between in the interval $[-2,0]$, thus finding consistency between
our analysis and the theoretical prediction eq.~(\ref{eq:an_lb}).

\section{Comparison of $\Lambda$CDM and massive gravity models}
\label{app:RR}

\begin{figure}
  \begin{center}
    \includegraphics[width=.48\linewidth]{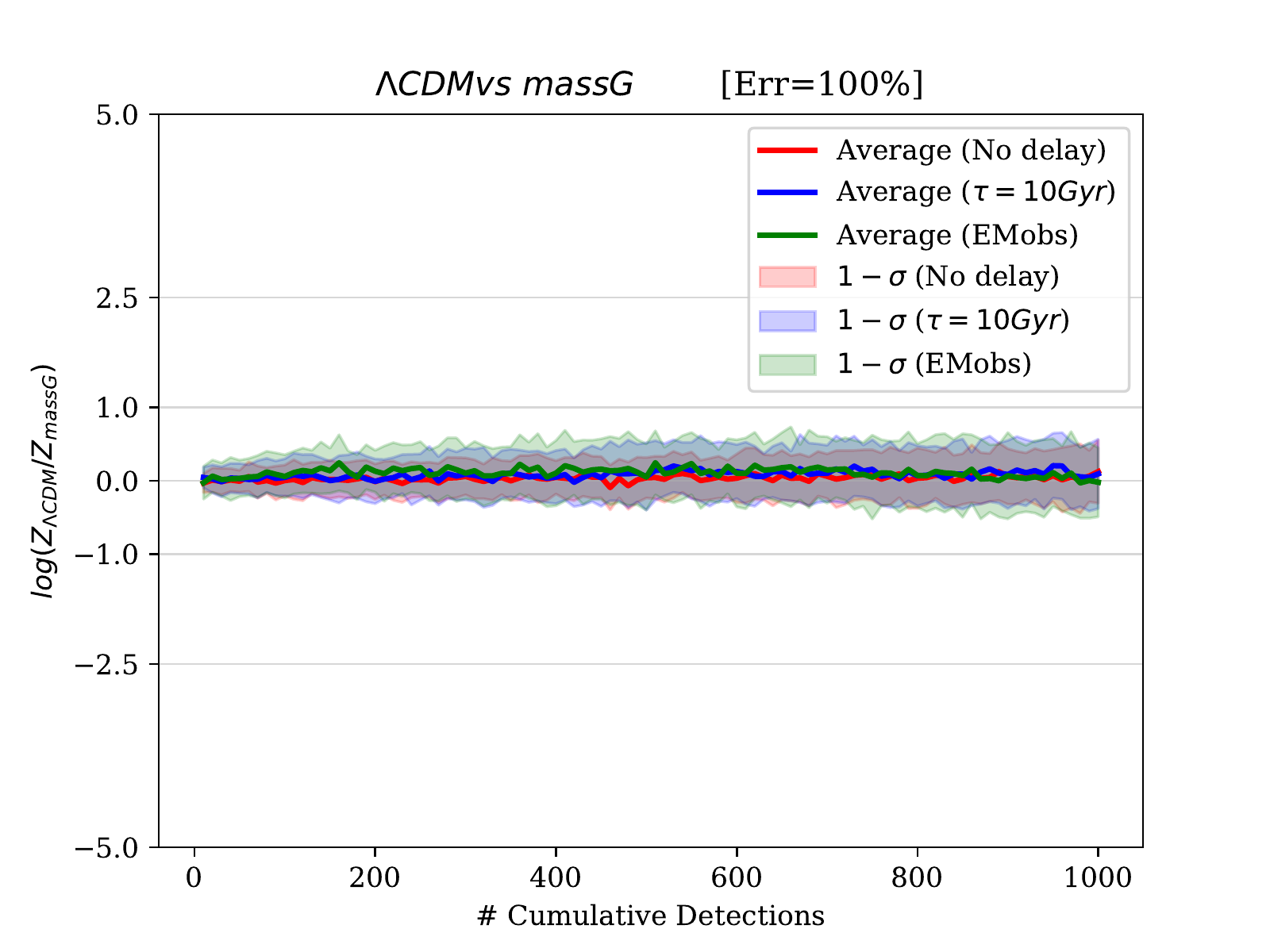}
    \includegraphics[width=.48\linewidth]{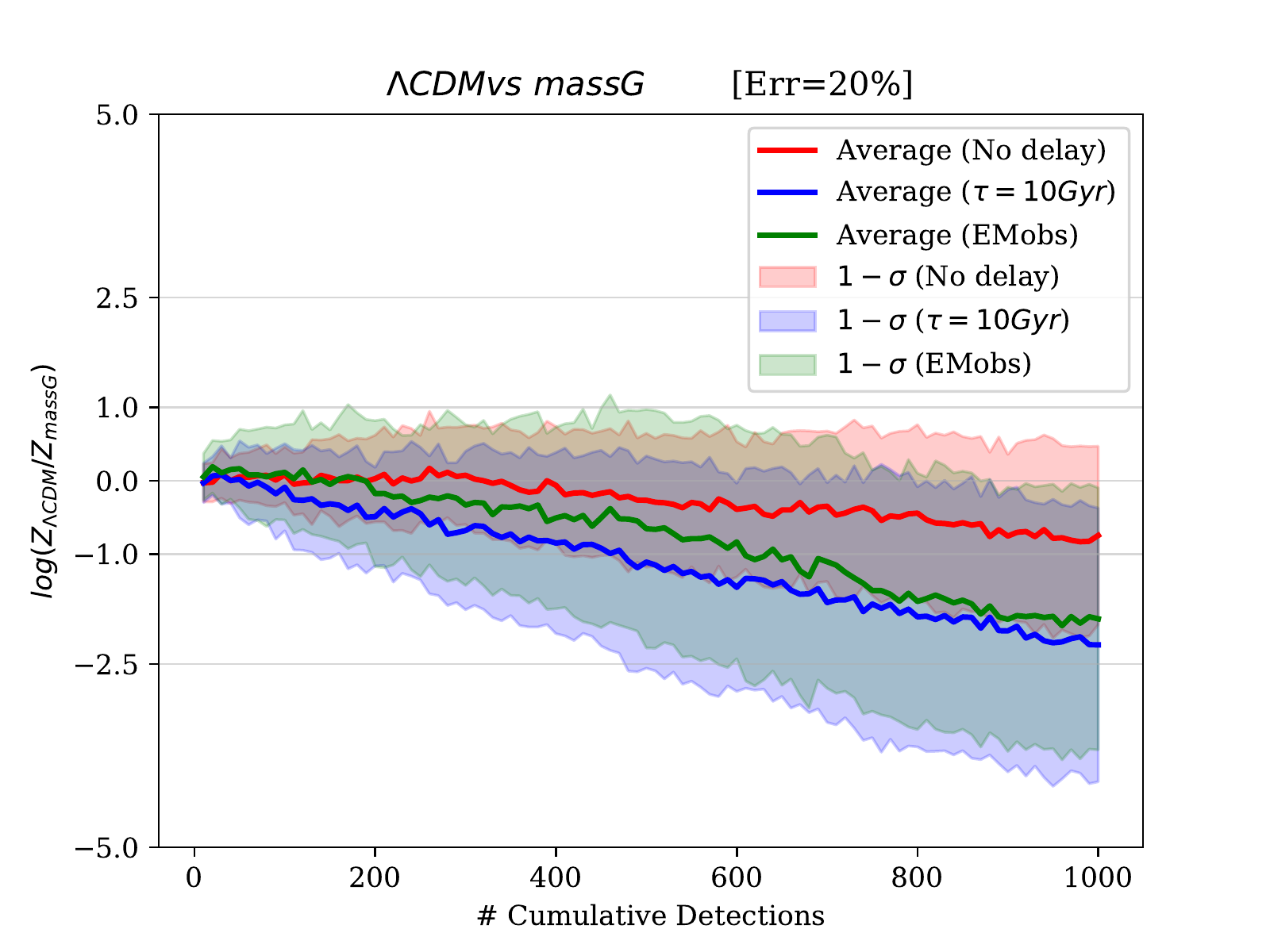}\\
    \caption{Evidence for $\Lambda$CDM versus \emph{massG} models with $1000$ events
      simulated according to the \emph{massG} model
      with standard uncertainty in the luminosity distance given by eq.~(\ref{eq:errdL}) (left) and uncertainty reduced to $20\%$ (right), 
      with merger rate equal to star formation rate, convolved with a Poisson-distributed $\tau=10$ Gyr delay as in eq.~(\ref{eq:poi_delay})
      between formation and merger and according to eq.~(\ref{eq:rateobs}).
      From 50 different injections realisations the shady regions corresponds to 1-$\sigma$ level around the thick line showing the average value.}
    \label{fig:LvsRRMassGRinjs}
\end{center}
\end{figure}

As a consistency check in fig.~\ref{fig:LvsRRMassGRinjs} we report the logarithm
of evidence ratios between $\Lambda$CDM and \emph{massG} models for \emph{massG} injections,
which indicates a better discriminating
power for simulations with reduced errors and distributed at relatively lower redshift.

In \cite{Belgacem:2018lbp} a comparison between the \emph{massG} and the
  $\Lambda$CDM model was performed by comparing respective $\chi^2$ values at
  best fit, whereas we report evidence ratios
  (which is roughly the likelihood integrated over search parameters with appropriate priors) which are described in sec.~\ref{sec:method}.
  Our results are broadly in agreement with those in \cite{Belgacem:2018lbp},
  which we could reproduce computing $\chi^2$ values (not reported here) instead
  of evidences.

\begin{thebibliography}{10}
\expandafter\ifx\csname url\endcsname\relax
  \def\url#1{\texttt{#1}}\fi
\expandafter\ifx\csname urlprefix\endcsname\relax\def\urlprefix{URL }\fi
\expandafter\ifx\csname href\endcsname\relax
  \def\href#1#2{#2} \def\path#1{#1}\fi

\bibitem{TheLIGOScientific:2017qsa}
B.~Abbott, et~al., {GW170817: Observation of Gravitational Waves from a Binary
  Neutron Star Inspiral}, Phys. Rev. Lett. 119~(16) (2017) 161101.
\newblock \href {http://arxiv.org/abs/1710.05832} {\path{arXiv:1710.05832}},
  \href {http://dx.doi.org/10.1103/PhysRevLett.119.161101}
  {\path{doi:10.1103/PhysRevLett.119.161101}}.

\bibitem{GBM:2017lvd}
B.~P. Abbott, et~al., {Multi-messenger Observations of a Binary Neutron Star
  Merger}, Astrophys. J. 848~(2) (2017) L12.
\newblock \href {http://arxiv.org/abs/1710.05833} {\path{arXiv:1710.05833}},
  \href {http://dx.doi.org/10.3847/2041-8213/aa91c9}
  {\path{doi:10.3847/2041-8213/aa91c9}}.

\bibitem{Schutz:1986gp}
B.~F. Schutz, {Determining the Hubble Constant from Gravitational Wave
  Observations}, Nature 323 (1986) 310--311.
\newblock \href {http://dx.doi.org/10.1038/323310a0}
  {\path{doi:10.1038/323310a0}}.

\bibitem{Holz:2005df}
D.~E. Holz, S.~A. Hughes, {Using gravitational-wave standard sirens},
  Astrophys. J. 629 (2005) 15--22.
\newblock \href {http://arxiv.org/abs/astro-ph/0504616}
  {\path{arXiv:astro-ph/0504616}}, \href {http://dx.doi.org/10.1086/431341}
  {\path{doi:10.1086/431341}}.

\bibitem{Riess:2016jrr}
A.~G. Riess, et~al., {A 2.4\% Determination of the Local Value of the Hubble
  Constant}, Astrophys. J. 826~(1) (2016) 56.
\newblock \href {http://arxiv.org/abs/1604.01424} {\path{arXiv:1604.01424}},
  \href {http://dx.doi.org/10.3847/0004-637X/826/1/56}
  {\path{doi:10.3847/0004-637X/826/1/56}}.

\bibitem{Messenger:2011gi}
C.~Messenger, J.~Read, {Measuring a cosmological distance-redshift relationship
  using only gravitational wave observations of binary neutron star
  coalescences}, Phys. Rev. Lett. 108 (2012) 091101.
\newblock \href {http://arxiv.org/abs/1107.5725} {\path{arXiv:1107.5725}},
  \href {http://dx.doi.org/10.1103/PhysRevLett.108.091101}
  {\path{doi:10.1103/PhysRevLett.108.091101}}.

\bibitem{DelPozzo:2011yh}
W.~Del~Pozzo, {Inference of the cosmological parameters from gravitational
  waves: application to second generation interferometers}, Phys. Rev. D86
  (2012) 043011.
\newblock \href {http://arxiv.org/abs/1108.1317} {\path{arXiv:1108.1317}},
  \href {http://dx.doi.org/10.1103/PhysRevD.86.043011}
  {\path{doi:10.1103/PhysRevD.86.043011}}.

\bibitem{Punturo:2010zz}
M.~Punturo, et~al., {The Einstein Telescope: A third-generation gravitational
  wave observatory}, Class. Quant. Grav. 27 (2010) 194002.
\newblock \href {http://dx.doi.org/10.1088/0264-9381/27/19/194002}
  {\path{doi:10.1088/0264-9381/27/19/194002}}.

\bibitem{Evans:2016mbw}
B.~P. Abbott, et~al., {Exploring the Sensitivity of Next Generation
  Gravitational Wave Detectors}, Class. Quant. Grav. 34~(4) (2017) 044001.
\newblock \href {http://arxiv.org/abs/1607.08697} {\path{arXiv:1607.08697}},
  \href {http://dx.doi.org/10.1088/1361-6382/aa51f4}
  {\path{doi:10.1088/1361-6382/aa51f4}}.

\bibitem{Harry:2010zz}
G.~M. Harry, {Advanced LIGO: The next generation of gravitational wave
  detectors}, Class. Quant. Grav. 27 (2010) 084006.
\newblock \href {http://dx.doi.org/10.1088/0264-9381/27/8/084006}
  {\path{doi:10.1088/0264-9381/27/8/084006}}.

\bibitem{TheVirgo:2014hva}
F.~Acernese, et~al., {Advanced Virgo: a second-generation interferometric
  gravitational wave detector}, Class. Quant. Grav. 32~(2) (2015) 024001.
\newblock \href {http://arxiv.org/abs/1408.3978} {\path{arXiv:1408.3978}},
  \href {http://dx.doi.org/10.1088/0264-9381/32/2/024001}
  {\path{doi:10.1088/0264-9381/32/2/024001}}.

\bibitem{Belgacem:2019tbw}
E.~Belgacem, Y.~Dirian, S.~Foffa, E.~J. Howell, M.~Maggiore, T.~Regimbau,
  {Cosmology and dark energy from joint gravitational wave-GRB observations},
  JCAP 1908 (2019) 015.
\newblock \href {http://arxiv.org/abs/1907.01487} {\path{arXiv:1907.01487}},
  \href {http://dx.doi.org/10.1088/1475-7516/2019/08/015}
  {\path{doi:10.1088/1475-7516/2019/08/015}}.

\bibitem{Riess_last}
A.~G. Riess, S.~Casertano, W.~Yuan, L.~Macri, J.~Anderson, J.~W. MacKenty,
  J.~B. Bowers, K.~I. Clubb, A.~V. Filippenko, D.~O. Jones, B.~E. Tucker, New
  parallaxes of galactic cepheids from spatially scanning the hubble space
  telescope: Implications for the hubble constant, Astrophys. J. 136 (2018)
  855.
\newblock \href {http://arxiv.org/abs/1801.01120} {\path{arXiv:1801.01120}}.

\bibitem{Aghanim:2018eyx}
N.~Aghanim, et~al., {Planck 2018 results. VI. Cosmological parameters}, Astron.
  Astrophys. 641 (2020) A6.
\newblock \href {http://arxiv.org/abs/1807.06209} {\path{arXiv:1807.06209}},
  \href {http://dx.doi.org/10.1051/0004-6361/201833910}
  {\path{doi:10.1051/0004-6361/201833910}}.

\bibitem{Birrer:2018vtm}
S.~Birrer, et~al., {H0LiCOW - IX. Cosmographic analysis of the doubly imaged
  quasar SDSS 1206+4332 and a new measurement of the Hubble constant}, Mon.
  Not. Roy. Astron. Soc. 484 (2019) 4726.
\newblock \href {http://arxiv.org/abs/1809.01274} {\path{arXiv:1809.01274}},
  \href {http://dx.doi.org/10.1093/mnras/stz200}
  {\path{doi:10.1093/mnras/stz200}}.

\bibitem{refId0}
{Khetan, Nandita}, {Izzo, Luca}, {Branchesi, Marica}, {Wojtak, Radoslaw},
  {Cantiello, Michele}, {Murugeshan, Chandrashekar}, {Agnello, Adriano},
  {Cappellaro, Enrico}, {Della Valle, Massimo}, {Gall, Christa}, {Hjorth,
  Jens}, {Benetti, Stefano}, {Brocato, Enzo}, {Burke, Jamison}, {Hiramatsu,
  Daichi}, {Howell, D. Andrew}, {Tomasella, Lina}, {Valenti, Stefano},
  \href{https://doi.org/10.1051/0004-6361/202039196}{A new measurement of the
  hubble constant using type ia supernovae calibrated with surface brightness
  fluctuations}, A\&A 647 (2021) A72.
\newblock \href {http://dx.doi.org/10.1051/0004-6361/202039196}
  {\path{doi:10.1051/0004-6361/202039196}}.
\newline\urlprefix\url{https://doi.org/10.1051/0004-6361/202039196}

\bibitem{Verde:2019ivm}
L.~Verde, T.~Treu, A.~G. Riess, {Tensions between the Early and the Late
  Universe}, Nature Astron. 3 (2019) 891.
\newblock \href {http://arxiv.org/abs/1907.10625} {\path{arXiv:1907.10625}},
  \href {http://dx.doi.org/10.1038/s41550-019-0902-0}
  {\path{doi:10.1038/s41550-019-0902-0}}.

\bibitem{Dalal:2006qt}
N.~Dalal, D.~E. Holz, S.~A. Hughes, B.~Jain, {Short grb and binary black hole
  standard sirens as a probe of dark energy}, Phys. Rev. D74 (2006) 063006.
\newblock \href {http://arxiv.org/abs/astro-ph/0601275}
  {\path{arXiv:astro-ph/0601275}}, \href
  {http://dx.doi.org/10.1103/PhysRevD.74.063006}
  {\path{doi:10.1103/PhysRevD.74.063006}}.

\bibitem{Sathyaprakash:2009xt}
B.~S. Sathyaprakash, B.~F. Schutz, C.~Van Den~Broeck, {Cosmography with the
  Einstein Telescope}, Class. Quant. Grav. 27 (2010) 215006.
\newblock \href {http://arxiv.org/abs/0906.4151} {\path{arXiv:0906.4151}},
  \href {http://dx.doi.org/10.1088/0264-9381/27/21/215006}
  {\path{doi:10.1088/0264-9381/27/21/215006}}.

\bibitem{Cai:2016sby}
R.-G. Cai, T.~Yang, {Estimating cosmological parameters by the simulated data
  of gravitational waves from the Einstein Telescope}, Phys. Rev. D95~(4)
  (2017) 044024.
\newblock \href {http://arxiv.org/abs/1608.08008} {\path{arXiv:1608.08008}},
  \href {http://dx.doi.org/10.1103/PhysRevD.95.044024}
  {\path{doi:10.1103/PhysRevD.95.044024}}.

\bibitem{Zhang:2018byx}
X.-N. Zhang, L.-F. Wang, J.-F. Zhang, X.~Zhang, {Improving cosmological
  parameter estimation with the future gravitational-wave standard siren
  observation from the Einstein Telescope}, Phys. Rev. D99~(6) (2019) 063510.
\newblock \href {http://arxiv.org/abs/1804.08379} {\path{arXiv:1804.08379}},
  \href {http://dx.doi.org/10.1103/PhysRevD.99.063510}
  {\path{doi:10.1103/PhysRevD.99.063510}}.

\bibitem{Chen:2017rfc}
H.-Y. Chen, M.~Fishbach, D.~E. Holz, {A two per cent Hubble constant
  measurement from standard sirens within five years}, Nature 562~(7728) (2018)
  545--547.
\newblock \href {http://arxiv.org/abs/1712.06531} {\path{arXiv:1712.06531}},
  \href {http://dx.doi.org/10.1038/s41586-018-0606-0}
  {\path{doi:10.1038/s41586-018-0606-0}}.

\bibitem{Camera:2013xfa}
S.~Camera, A.~Nishizawa, {Beyond Concordance Cosmology with Magnification of
  Gravitational-Wave Standard Sirens}, Phys. Rev. Lett. 110~(15) (2013) 151103.
\newblock \href {http://arxiv.org/abs/1303.5446} {\path{arXiv:1303.5446}},
  \href {http://dx.doi.org/10.1103/PhysRevLett.110.151103}
  {\path{doi:10.1103/PhysRevLett.110.151103}}.

\bibitem{Nishizawa:2017nef}
A.~Nishizawa, {Generalized framework for testing gravity with
  gravitational-wave propagation. I. Formulation}, Phys. Rev. D97~(10) (2018)
  104037.
\newblock \href {http://arxiv.org/abs/1710.04825} {\path{arXiv:1710.04825}},
  \href {http://dx.doi.org/10.1103/PhysRevD.97.104037}
  {\path{doi:10.1103/PhysRevD.97.104037}}.

\bibitem{Arai:2017hxj}
S.~Arai, A.~Nishizawa, {Generalized framework for testing gravity with
  gravitational-wave propagation. II. Constraints on Horndeski theory}, Phys.
  Rev. D97~(10) (2018) 104038.
\newblock \href {http://arxiv.org/abs/1711.03776} {\path{arXiv:1711.03776}},
  \href {http://dx.doi.org/10.1103/PhysRevD.97.104038}
  {\path{doi:10.1103/PhysRevD.97.104038}}.

\bibitem{Nishizawa:2019rra}
A.~Nishizawa, S.~Arai, {Generalized framework for testing gravity with
  gravitational-wave propagation. III. Future prospect}, Phys. Rev. D99~(10)
  (2019) 104038.
\newblock \href {http://arxiv.org/abs/1901.08249} {\path{arXiv:1901.08249}},
  \href {http://dx.doi.org/10.1103/PhysRevD.99.104038}
  {\path{doi:10.1103/PhysRevD.99.104038}}.

\bibitem{Cai:2017yww}
R.-G. Cai, N.~Tamanini, T.~Yang, {Reconstructing the dark sector interaction
  with LISA}, JCAP 1705~(05) (2017) 031.
\newblock \href {http://arxiv.org/abs/1703.07323} {\path{arXiv:1703.07323}},
  \href {http://dx.doi.org/10.1088/1475-7516/2017/05/031}
  {\path{doi:10.1088/1475-7516/2017/05/031}}.

\bibitem{Abbott:2018wog}
T.~M.~C. Abbott, et~al., {First Cosmology Results using Type Ia Supernovae from
  the Dark Energy Survey: Constraints on Cosmological Parameters}, Astrophys.
  J. 872~(2) (2019) L30.
\newblock \href {http://arxiv.org/abs/1811.02374} {\path{arXiv:1811.02374}},
  \href {http://dx.doi.org/10.3847/2041-8213/ab04fa}
  {\path{doi:10.3847/2041-8213/ab04fa}}.

\bibitem{Taylor:2012db}
S.~R. Taylor, J.~R. Gair, {Cosmology with the lights off: standard sirens in
  the Einstein Telescope era}, Phys. Rev. D86 (2012) 023502.
\newblock \href {http://arxiv.org/abs/1204.6739} {\path{arXiv:1204.6739}},
  \href {http://dx.doi.org/10.1103/PhysRevD.86.023502}
  {\path{doi:10.1103/PhysRevD.86.023502}}.

\bibitem{Blanchard:2019oqi}
A.~Blanchard, et~al., {Euclid preparation: VII. Forecast validation for Euclid
  cosmological probes}, Astron. Astrophys. 642 (2020) A191.
\newblock \href {http://arxiv.org/abs/1910.09273} {\path{arXiv:1910.09273}},
  \href {http://dx.doi.org/10.1051/0004-6361/202038071}
  {\path{doi:10.1051/0004-6361/202038071}}.

\bibitem{Chevallier:2000qy}
M.~Chevallier, D.~Polarski, {Accelerating universes with scaling dark matter},
  Int. J. Mod. Phys. D10 (2001) 213--224.
\newblock \href {http://arxiv.org/abs/gr-qc/0009008}
  {\path{arXiv:gr-qc/0009008}}, \href
  {http://dx.doi.org/10.1142/S0218271801000822}
  {\path{doi:10.1142/S0218271801000822}}.

\bibitem{Linder:2002et}
E.~V. Linder, {Exploring the expansion history of the universe}, Phys. Rev.
  Lett. 90 (2003) 091301.
\newblock \href {http://arxiv.org/abs/astro-ph/0208512}
  {\path{arXiv:astro-ph/0208512}}, \href
  {http://dx.doi.org/10.1103/PhysRevLett.90.091301}
  {\path{doi:10.1103/PhysRevLett.90.091301}}.

\bibitem{Belgacem:2017cqo}
E.~Belgacem, Y.~Dirian, S.~Foffa, M.~Maggiore, {Nonlocal gravity. Conceptual
  aspects and cosmological predictions}, JCAP 1803~(03) (2018) 002.
\newblock \href {http://arxiv.org/abs/1712.07066} {\path{arXiv:1712.07066}},
  \href {http://dx.doi.org/10.1088/1475-7516/2018/03/002}
  {\path{doi:10.1088/1475-7516/2018/03/002}}.

\bibitem{Belgacem:2018lbp}
E.~Belgacem, Y.~Dirian, S.~Foffa, M.~Maggiore, {Modified gravitational-wave
  propagation and standard sirens}, Phys. Rev. D98~(2) (2018) 023510.
\newblock \href {http://arxiv.org/abs/1805.08731} {\path{arXiv:1805.08731}},
  \href {http://dx.doi.org/10.1103/PhysRevD.98.023510}
  {\path{doi:10.1103/PhysRevD.98.023510}}.

\bibitem{Trotta:2008qt}
R.~Trotta, {Bayes in the sky: Bayesian inference and model selection in
  cosmology}, Contemp. Phys. 49 (2008) 71--104.
\newblock \href {http://arxiv.org/abs/0803.4089} {\path{arXiv:0803.4089}},
  \href {http://dx.doi.org/10.1080/00107510802066753}
  {\path{doi:10.1080/00107510802066753}}.

\bibitem{Zhao:2010sz}
W.~Zhao, C.~Van Den~Broeck, D.~Baskaran, T.~G.~F. Li, {Determination of Dark
  Energy by the Einstein Telescope: Comparing with CMB, BAO and SNIa
  Observations}, Phys. Rev. D83 (2011) 023005.
\newblock \href {http://arxiv.org/abs/1009.0206} {\path{arXiv:1009.0206}},
  \href {http://dx.doi.org/10.1103/PhysRevD.83.023005}
  {\path{doi:10.1103/PhysRevD.83.023005}}.

\bibitem{Vitale:2016icu}
S.~Vitale, M.~Evans, {Parameter estimation for binary black holes with networks
  of third generation gravitational-wave detectors}, Phys. Rev. D95~(6) (2017)
  064052.
\newblock \href {http://arxiv.org/abs/1610.06917} {\path{arXiv:1610.06917}},
  \href {http://dx.doi.org/10.1103/PhysRevD.95.064052}
  {\path{doi:10.1103/PhysRevD.95.064052}}.

\bibitem{Sathyaprakash:2019rom}
B.~S. Sathyaprakash, et~al., {Multimessenger Universe with Gravitational Waves
  from Binaries}\href {http://arxiv.org/abs/1903.09277}
  {\path{arXiv:1903.09277}}.

\bibitem{Madau:2014bja}
P.~Madau, M.~Dickinson, {Cosmic Star Formation History}, Ann. Rev. Astron.
  Astrophys. 52 (2014) 415--486.
\newblock \href {http://arxiv.org/abs/1403.0007} {\path{arXiv:1403.0007}},
  \href {http://dx.doi.org/10.1146/annurev-astro-081811-125615}
  {\path{doi:10.1146/annurev-astro-081811-125615}}.

\bibitem{Vitale:2018yhm}
S.~Vitale, W.~M. Farr, K.~Ng, C.~L. Rodriguez, {Measuring the star formation
  rate with gravitational waves from binary black holes}, Astrophys. J. Lett.
  886~(1) (2019) L1.
\newblock \href {http://arxiv.org/abs/1808.00901} {\path{arXiv:1808.00901}},
  \href {http://dx.doi.org/10.3847/2041-8213/ab50c0}
  {\path{doi:10.3847/2041-8213/ab50c0}}.

\bibitem{Safarzadeh:2019pis}
M.~Safarzadeh, E.~Berger, K.~K.~Y. Ng, H.-Y. Chen, S.~Vitale, C.~Whittle,
  E.~Scannapieco, {Measuring the delay time distribution of binary neutron
  stars. II. Using the redshift distribution from third-generation
  gravitational wave detectors network}, Astrophys. J. 878~(1) (2019) L13.
\newblock \href {http://arxiv.org/abs/1904.10976} {\path{arXiv:1904.10976}},
  \href {http://dx.doi.org/10.3847/2041-8213/ab22be}
  {\path{doi:10.3847/2041-8213/ab22be}}.

\bibitem{Scolnic:2017caz}
D.~M. Scolnic, et~al., {The Complete Light-curve Sample of Spectroscopically
  Confirmed SNe Ia from Pan-STARRS1 and Cosmological Constraints from the
  Combined Pantheon Sample}, Astrophys. J. 859~(2) (2018) 101.
\newblock \href {http://arxiv.org/abs/1710.00845} {\path{arXiv:1710.00845}},
  \href {http://dx.doi.org/10.3847/1538-4357/aab9bb}
  {\path{doi:10.3847/1538-4357/aab9bb}}.

\bibitem{Meacher:2015iua}
D.~Meacher, M.~Coughlin, S.~Morris, T.~Regimbau, N.~Christensen, S.~Kandhasamy,
  V.~Mandic, J.~D. Romano, E.~Thrane, {Mock data and science challenge for
  detecting an astrophysical stochastic gravitational-wave background with
  Advanced LIGO and Advanced Virgo}, Phys. Rev. D92~(6) (2015) 063002.
\newblock \href {http://arxiv.org/abs/1506.06744} {\path{arXiv:1506.06744}},
  \href {http://dx.doi.org/10.1103/PhysRevD.92.063002}
  {\path{doi:10.1103/PhysRevD.92.063002}}.

\bibitem{Abbott:2019yzh}
B.~P. Abbott, et~al., {A Gravitational-wave Measurement of the Hubble Constant
  Following the Second Observing Run of Advanced LIGO and Virgo}, Astrophys. J.
  909~(2) (2021) 218.
\newblock \href {http://arxiv.org/abs/1908.06060} {\path{arXiv:1908.06060}},
  \href {http://dx.doi.org/10.3847/1538-4357/abdcb7}
  {\path{doi:10.3847/1538-4357/abdcb7}}.

\bibitem{Bonvin:2005ps}
C.~Bonvin, R.~Durrer, M.~A. Gasparini, {Fluctuations of the luminosity
  distance}, Phys. Rev. D73 (2006) 023523, [Erratum: Phys.
  Rev.D85,029901(2012)].
\newblock \href {http://arxiv.org/abs/astro-ph/0511183}
  {\path{arXiv:astro-ph/0511183}}, \href
  {http://dx.doi.org/10.1103/PhysRevD.85.029901, 10.1103/PhysRevD.73.023523}
  {\path{doi:10.1103/PhysRevD.85.029901, 10.1103/PhysRevD.73.023523}}.

\bibitem{Hirata:2010ba}
C.~M. Hirata, D.~E. Holz, C.~Cutler, {Reducing the weak lensing noise for the
  gravitational wave Hubble diagram using the non-Gaussianity of the
  magnification distribution}, Phys. Rev. D81 (2010) 124046.
\newblock \href {http://arxiv.org/abs/1004.3988} {\path{arXiv:1004.3988}},
  \href {http://dx.doi.org/10.1103/PhysRevD.81.124046}
  {\path{doi:10.1103/PhysRevD.81.124046}}.

\bibitem{Bonvin:2016qxr}
C.~Bonvin, C.~Caprini, R.~Sturani, N.~Tamanini, {Effect of matter structure on
  the gravitational waveform}, Phys. Rev. D95~(4) (2017) 044029.
\newblock \href {http://arxiv.org/abs/1609.08093} {\path{arXiv:1609.08093}},
  \href {http://dx.doi.org/10.1103/PhysRevD.95.044029}
  {\path{doi:10.1103/PhysRevD.95.044029}}.

\bibitem{Gordon:2007zw}
C.~Gordon, K.~Land, A.~Slosar, {Cosmological Constraints from Type Ia
  Supernovae Peculiar Velocity Measurements}, Phys. Rev. Lett. 99 (2007)
  081301.
\newblock \href {http://arxiv.org/abs/0705.1718} {\path{arXiv:0705.1718}},
  \href {http://dx.doi.org/10.1103/PhysRevLett.99.081301}
  {\path{doi:10.1103/PhysRevLett.99.081301}}.

\bibitem{Vitale:2018nif}
S.~Vitale, C.~Whittle, {Characterization of binary black holes by heterogeneous
  gravitational-wave networks}, Phys. Rev. D98~(2) (2018) 024029.
\newblock \href {http://arxiv.org/abs/1804.07866} {\path{arXiv:1804.07866}},
  \href {http://dx.doi.org/10.1103/PhysRevD.98.024029}
  {\path{doi:10.1103/PhysRevD.98.024029}}.

\bibitem{Tamanini:2016zlh}
N.~Tamanini, C.~Caprini, E.~Barausse, A.~Sesana, A.~Klein, A.~Petiteau,
  {Science with the space-based interferometer eLISA. III: Probing the
  expansion of the Universe using gravitational wave standard sirens}, JCAP
  1604~(04) (2016) 002.
\newblock \href {http://arxiv.org/abs/1601.07112} {\path{arXiv:1601.07112}},
  \href {http://dx.doi.org/10.1088/1475-7516/2016/04/002}
  {\path{doi:10.1088/1475-7516/2016/04/002}}.

\bibitem{Dirian:2017pwp}
Y.~Dirian, {Changing the Bayesian prior: Absolute neutrino mass constraints in
  nonlocal gravity}, Phys. Rev. D96~(8) (2017) 083513.
\newblock \href {http://arxiv.org/abs/1704.04075} {\path{arXiv:1704.04075}},
  \href {http://dx.doi.org/10.1103/PhysRevD.96.083513}
  {\path{doi:10.1103/PhysRevD.96.083513}}.

\bibitem{Mukherjee:2005wg}
P.~Mukherjee, D.~Parkinson, A.~R. Liddle, {A nested sampling algorithm for
  cosmological model selection}, Astrophys. J. 638 (2006) L51--L54.
\newblock \href {http://arxiv.org/abs/astro-ph/0508461}
  {\path{arXiv:astro-ph/0508461}}, \href {http://dx.doi.org/10.1086/501068}
  {\path{doi:10.1086/501068}}.

\bibitem{Skilling:2006gxv}
J.~Skilling, {Nested sampling for general Bayesian computation}, Bayesian
  Analysis 1~(4) (2006) 833--859.
\newblock \href {http://dx.doi.org/10.1214/06-BA127}
  {\path{doi:10.1214/06-BA127}}.

\bibitem{jeffreys}
H.~Jeffreys, Theory of probability, Clarendon Press, Oxford, 1961, 3rd edn
  1998.

\bibitem{Speri:2020hwc}
L.~Speri, N.~Tamanini, R.~R. Caldwell, J.~R. Gair, B.~Wang, {Testing the Quasar
  Hubble Diagram with LISA Standard Sirens}\href
  {http://arxiv.org/abs/2010.09049} {\path{arXiv:2010.09049}}.

\bibitem{Risaliti:2018reu}
G.~Risaliti, E.~Lusso, {Cosmological constraints from the Hubble diagram of
  quasars at high redshifts}, Nature Astron. 3~(3) (2019) 272--277.
\newblock \href {http://arxiv.org/abs/1811.02590} {\path{arXiv:1811.02590}},
  \href {http://dx.doi.org/10.1038/s41550-018-0657-z}
  {\path{doi:10.1038/s41550-018-0657-z}}.

\end{thebibliography}
\end{document}